\documentclass[onecolumn,superscriptaddress,11pt,letterpaper,accepted=2020-10-29]{quantumarticle}
\pdfoutput=1
\usepackage{amsmath, amsthm, amssymb,graphicx,color}
\usepackage[numbers,sort&compress]{natbib}
\usepackage[utf8]{inputenc}
\usepackage{ragged2e}
\usepackage{hyperref}
\usepackage{xspace}
\usepackage{url}

\usepackage{wrapfig}
\hypersetup{colorlinks=true, pdftitle=A volumetric framework for quantum computer benchmarks, citecolor=blue, linkcolor=blue}

\begin{document}

\newcommand{\rrangle}{\rangle\!\rangle} \newcommand{\llangle}{\langle\!\langle}

\newcommand{\cM}{\mathcal{M}}
\newcommand{\cP}{\mathcal{P}}
\newcommand{\cG}{\mathcal{G}}
\newcommand{\cD}{\mathcal{D}}
\newcommand{\cE}{\mathcal{E}}
\newcommand{\cL}{\mathcal{L}}
\newcommand{\cU}{\mathcal{U}}
\newcommand{\cH}{\mathcal{H}}

\newcommand{\reals}{\mathbb{R}}
\newcommand{\complex}{\mathbb{C}}

\newcommand{\expect}[1]{\ensuremath{\left\langle#1\right\rangle}}

\newcommand{\qhat}{\widehat{q}}
\newcommand{\rhohat}{\widehat{\rho}}
\newcommand{\rhoMLE}{\widehat{\rho}_{\mathrm{MLE}}}
\newcommand{\rhotrue}{\rho_{\mathrm{true}}}
\newcommand{\Rhat}{\widehat{\mathcal{R}}}
\newcommand{\Vbar}{\overline{V}}

\newcommand{\MS}{M\o lmer-S\o rensen\xspace}
\newcommand{\CNOT}{\ensuremath{\mathsf{CNOT}}\xspace}
\newcommand{\iSWAP}{\ensuremath{\mathsf{iSWAP}}\xspace}

\newcommand{\ket}[1]{\ensuremath{\left|#1\right\rangle}}
\newcommand{\bra}[1]{\ensuremath{\left\langle#1\right|}}
\newcommand{\braket}[2]{\ensuremath{\left\langle#1|#2\right\rangle}}

\newcommand{\ketbra}[2]{\ket{#1}\!\!\bra{#2}}
\newcommand{\braopket}[3]{\ensuremath{\bra{#1}#2\ket{#3}}}
\newcommand{\proj}[1]{\ketbra{#1}{#1}}

\newcommand{\sket}[1]{\ensuremath{\left|#1\right\rrangle}}
\newcommand{\sbra}[1]{\ensuremath{\left\llangle#1\right|}}
\newcommand{\sbraket}[2]{\ensuremath{\left\llangle#1|#2\right\rrangle}}
\newcommand{\sketbra}[2]{\sket{#1}\!\!\sbra{#2}}
\newcommand{\sbraopket}[3]{\ensuremath{\sbra{#1}#2\sket{#3}}}
\newcommand{\sproj}[1]{\sketbra{#1}{#1}}
\newcommand{\order}[1]{\ensuremath{\mathcal{O}(#1)}}
\def\Id{1\!\mathrm{l}}
\newcommand{\Tr}{\mathrm{Tr}}
\newcommand{\Nparams}{N_{\mathrm{params}}}

\newcommand{\note}[1]{{\color{red}[#1]}}

\newcommand{\testsuite}{\ensuremath{\mathcal{C}}}
\newcommand{\kcy}[1]{\textcolor{violet}{[#1]\textsubscript{KCY}}}

\newcommand{\eg}{e.g.,\xspace}
\newcommand{\Eg}{E.g.,\xspace}
 
\makeatletter 
\renewcommand{\fnum@figure}{\textbf{Figure~\thefigure}}
\makeatother

\title{A volumetric framework for quantum computer benchmarks}

\author{Robin Blume-Kohout}
\affiliation{Quantum Performance Laboratory, Sandia National Laboratories\vspace{-.1cm} \\ Albuquerque, NM 87185 and Livermore, California 94550}
\author{Kevin Young}
\affiliation{Quantum Performance Laboratory, Sandia National Laboratories\vspace{-.1cm} \\ Albuquerque, NM 87185 and Livermore, California 94550}
\orcid{0000-0002-4679-4542}

\begin{abstract}
\noindent We propose a very large family of benchmarks for probing the performance of quantum computers. We call them \emph{volumetric benchmarks} (VBs) because they generalize IBM's benchmark for measuring quantum volume \cite{Cross18}. The quantum volume benchmark defines a family of \emph{square} circuits whose depth $d$ and width $w$ are the same.  A volumetric benchmark defines a family of \emph{rectangular} quantum circuits, for which $d$ and $w$ are uncoupled to allow the study of time/space performance trade-offs.  Each VB defines a mapping from circuit shapes -- $(w,d)$ pairs -- to test suites $\testsuite(w,d)$.  A test suite is an ensemble of test circuits that share a common structure. The test suite $\testsuite$ for a given circuit shape may be a single circuit $C$, a specific list of circuits $\{C_1\ldots C_N\}$ that must all be run, or a large set of possible circuits equipped with a distribution $Pr(C)$. The circuits in a given VB share a structure, which is limited only by designers' creativity. We list some known benchmarks, and other circuit families, that fit into the VB framework:  several families of random circuits, periodic circuits, and algorithm-inspired circuits. The last ingredient defining a benchmark is a success criterion that defines when a processor is judged to have ``passed'' a given test circuit.  We discuss several options.  Benchmark data can be analyzed in many ways to extract many properties, but we propose a simple, universal graphical summary of results that illustrates the Pareto frontier of the $d$ vs $w$ trade-off for the processor being benchmarked.

\noindent[1] A. Cross, et al., \emph{Phys. Rev. A}, 100, 032328, September 2019. \href{https://doi.org/10.1103/PhysRevA.100.032328}{DOI: 10.1103/PhysRevA.100.032328}.
\end{abstract}

\date{November 11, 2020}
\maketitle

\section{Introduction}
\label{sec:intro}
Quantum computing hardware is growing fast. Just a few years ago, two connected high-fidelity qubits were state of the art.  Now there are quantum processors with 5 \cite{BarendsNature14,DevittPRA16,DebnathNature16,LinkePNAS17}, 8 \cite{Wootton18}, 9 \cite{KellyNature15}, 11 \cite{IonQ18}, 16 \cite{Wootton18}, 19 \cite{Wootton18}, 20 \cite{IBMTokyo19}, even 53 \cite{AruteNature19,IBM19} controllable and interactable qubits. We need techniques for characterizing these devices, and assessing their performance. Much recent work has gone into QCVV (quantum characterization, verification, and validation), which focuses on measuring intrinsic noise/error behavior of 1- and 2-qubit gates. But with the emergence of larger processors that can be viewed and treated like baby quantum computers, there is increasing need for \emph{holistic benchmarks} that capture interesting aspects of processors' overall performance -- preferably at practically relevant tasks, whenever possible -- and enable comparing and contrasting different processors and architectures.

\begin{figure}[t!]
\centering
\includegraphics[width=.6\linewidth]{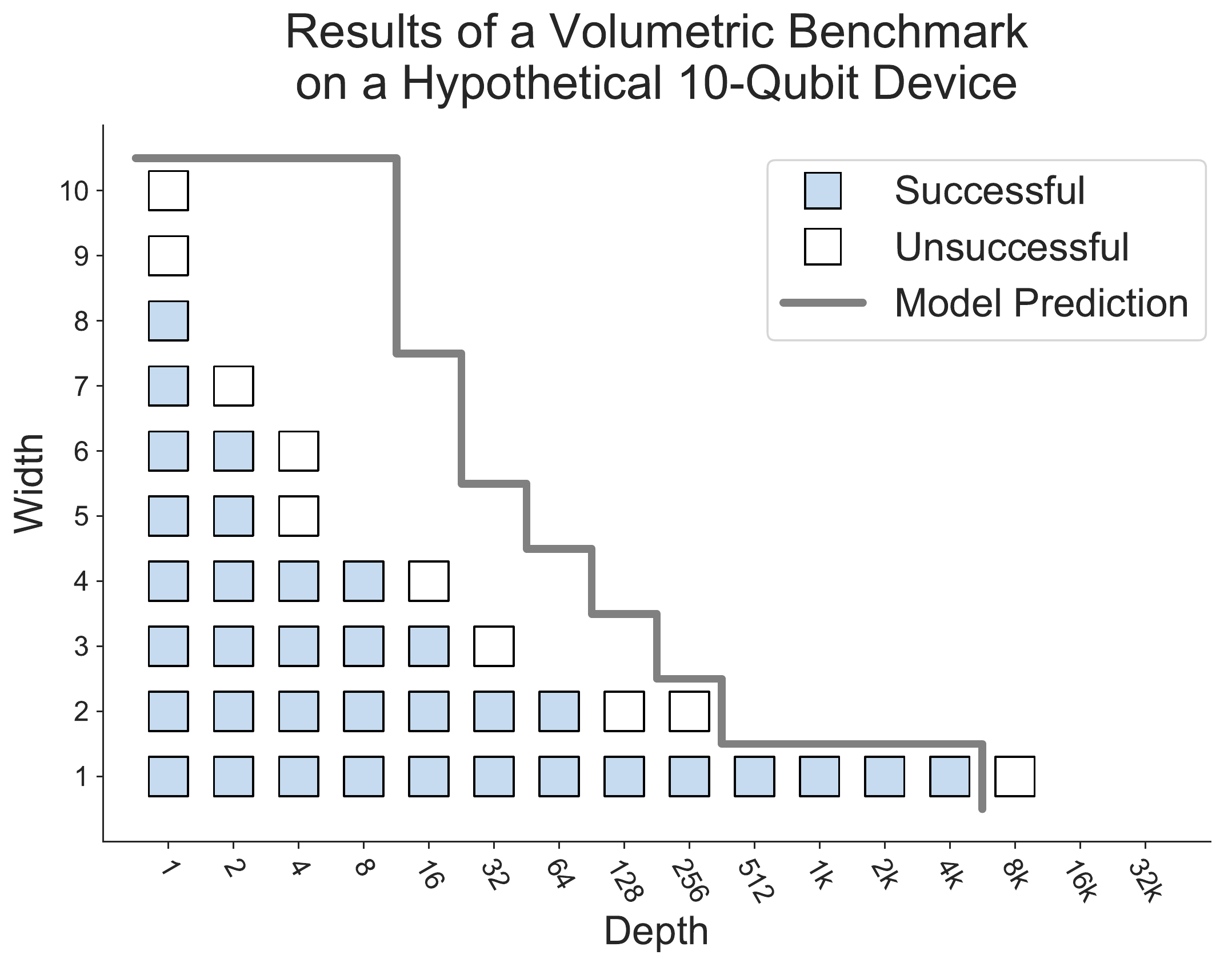}
\caption{Simple volumetric benchmarking plot for a 10-qubit device. The axes of the figure describe a range of possible circuit shapes (width/depth pairs). A solid mark is placed at a width/depth pair if a specified set of benchmarking circuits of that shape can be run successfully on the device.  An open mark indicates that the width/depth pair was tested, but failed the benchmark.  The solid line indicates a predicted pass/fail boundary based on a simple depolarizing model with error rates $5\times10^{-4}$ and $3\times10^{-3}$ for single- and two-qubit gates, respectively. More elaborate variations of this plotting style are presented in Sec.~\ref{sec:reporting_results}. \emph{This data is for illustration purposes only and does not come from a real device or rigorous simulation!}}
\label{fig:first_figure}
\end{figure}

IBM recently introduced a metric of quantum computer performance called ``quantum volume'' \cite{Cross18}. Quantum volume is intended to measure the size of a quantum processor's accessible state space. A processor with $n$ qubits has a $2^n$-dimensional state space. No matter how well it works, it can't access more than $2^n$ computational states. On the other hand, if the processor is poorly controlled, or suffers a lot of noise, then some of those $n$ qubits may be superfluous -- the full state space won't be accessible. Quantum volume aims to capture both of these limitations at once. It asks ``What's the largest number of qubits on which the processor can reliably produce a random state?'' Random states can be generated by random programs (circuits) \cite{BrandaoCMP16}, but the number of steps required (the \emph{depth} of the circuit) grows with the number of qubits. So demonstrating that a processor's quantum volume is at least $2^n$ requires demonstrating the ability to \emph{reliably} run a random quantum circuit with width (\# of qubits involved) and depth (\# of consecutive steps) \emph{both} at least $n$.  But while the quantum volume can provide a concise, high-level summary of system performance, very few quantum algorithms actually correspond to square circuits. Each aspect of a circuit's \emph{shape} (its width and depth) is independently interesting and can change dramatically depending on the target application. Probing precisely how a circuit's shape affects a given processor’s ability to run it successfully (see Fig.~\ref{fig:first_figure}) can serve as an important device diagnostic as well as an informative performance metric. 

IBM proposed a specific benchmark to probe quantum volume. By \emph{benchmark}, we mean a set of quantum circuits (a \emph{test suite}) together with instructions for how to run them (an \emph{experimental design}), an \emph{analysis procedure} for processing the raw results, and finally an \emph{interpretation rule} for drawing high-level conclusions. Benchmarks are often designed to measure one or more \emph{metrics}, and in this case the metric is quantum volume. Many other benchmarks intended to measure other specific metrics exist. Randomized benchmarking (RB) \cite{EmersonJOB05,EmersonScience07,KnillPRA08,MagesanPRL11,MagesanPRA12} is a benchmark designed to measure a certain fidelity metric. Long-sequence gate set tomography (GST) \cite{BlumeKohoutNC17,Greenbaum15,DehollainNJP16} -- a protocol that uses a tailored family of circuits to measure parameters of a gate set describing the processor's noisy operations -- defines a test suite that can be easily repurposed as a benchmark.  Many other protocols define test suites that can be used as benchmarks.

But benchmarks don't necessarily measure a well-defined metric. LINPACK \cite{Dongarra03} is a benchmark for classical computers that measures\ldots performance on LINPACK. That's a proxy for performance on general linear algebra, but it doesn't correspond to any intrinsic or pre-existing property of the computer. LINPACK effectively defines a new \emph{synthetic} metric. Most benchmarks for classical computers are like this. Some are intended to measure a well-defined metric like FLOPS, but even in those cases the benchmark defines a particular context (\eg linear algebra), and the metric's meaning varies across contexts.

Just what properties of quantum computers will be the most relevant -- which intrinsic and/or synthetic metrics will capture important aspects of performance -- is very unclear at this time. Quantum volume is intriguing, and a great first step, but almost certainly not the \emph{last} step in benchmarking quantum computers. In the absence of clear desiderata for metrics and/or benchmarks, we suggest that what the quantum computing community needs is a \emph{lot} of candidate benchmarks -- which may or may not be tied to intrinsic metrics -- that can be tested, deployed, applied, and made to compete against each other. The ones that prove the most useful will persist.

In this paper, we lay out a framework for a large and flexible family of benchmarks (see example in Fig.~\ref{fig:first_figure}). It was inspired by IBM's benchmark for quantum volume, so we call them \emph{volumetric benchmarks}. We give a few examples, mostly to demonstrate how the framework works, but these are not intended to be exhaustive! The most important property of a test circuit is its structure, because circuits with different structures can probe (or suppress) totally different noise/error properties. The volumetric framework allows circuit families with radically different structures, which we hope will allow the construction of volumetric benchmarks that capture a radically wide range of behaviors.

Like most ideas, this one is not entirely original.  Quantum volume is the most obvious and explicit precedent, and our explicit intent here is to generalize it.  More generally, it is a basic and well-known truth that both depth \emph{and} width of the circuits that can be run on a quantum processor are interesting.  Emerson and Wallman proposed a metric -- called \emph{quantum circuit capacity} in a 2017 talk \cite{Emerson2017} or, more recently, \emph{quantum processing power} \cite{Emerson2019p} -- based on a processor's ability to run $d \times w$ rectangular circuits derived from cycle benchmarking \cite{Erhard19}. Bishop, responding to a question in an online forum, suggested that the quantum volume could serve as an ``executive summary'' of device performance, but that one should additionally report benchmarking results at a range of width/depth pairs \cite{Bishop18}. In fact, almost every aspect of the framework we propose here is inspired by one or more proposed approaches to benchmarking.  Our main contributions are (1) their integration, and (2) the reporting and visualization framework in the last section of the paper.  We have attempted to cite prior work of which we're aware, and apologize in advance for any failures.

\section{Motivations for generalizing (why not just quantum volume?)}
\label{sec:motivation}

An obvious question is ``Is it even necessary to generalize beyond quantum volume, given that quantum volume already exists?''

Although quantum volume is an excellent (and inspiring) idea, neither quantum volume nor any other single metric will capture all the important information about holistic performance. Quantum volume measures the processor's ability to run ``square'' circuits of a particular form. While this correctly captures the idea that useful circuits' depth must increase with width (number of qubits), the circuits for most algorithms aren't square. Shor's factoring algorithm \cite{Shor94}, for example, requires $O(n)$ qubits but $O(n^3)$ depth. Grover's algorithm \cite{GroverPRL97} applied to search over $n$-bit strings requires $O(2^{n/2})$ repetitions of an oracle subroutine that, itself, generally has depth $O(\mathrm{poly}(n))$. On the other hand, many quantum algorithms can be reformulated to require only $\mathrm{polylog}(n)$ depth, at the cost of much larger width \cite{CleveFOCS00}.

The intuition behind quantum volume is that taking full advantage of $n$ qubits requires the ability to access any part of their Hilbert space, and that task -- also known as \emph{scrambling} the initial state \cite{SekinoJHEP08} -- demands $O(n)$ depth. There's significant truth in this, but it's incomplete. Useful quantum algorithms do not always explore the whole Hilbert space -- for example, Grover's algorithm operates within a 2-dimensional subspace spanned by the answer and the uniform superposition. Conversely, just one scrambling isn't usually enough; algorithms often need to follow a long and complex path through Hilbert space.

For all these reasons, users and designers of quantum computers will want to understand the trade-offs between width and depth. A 100-qubit processor that can perform 1000 layers of gates is more useful \emph{for certain tasks} than one that can just perform 100 layers. Similarly, there are situations where 1000 qubits that can perform 100 layers are more useful than just 100 qubits that can also perform 100 layers. Square circuits might be the \emph{most} important ones, but by no means the \emph{only} important circuits.

There's another important reason to design, run, and study other benchmarks besides the one that defines quantum volume. Different errors affect different circuits in different ways. Independent local depolarizing errors affect pretty much all circuits in the same predictable way. But unitary or ``coherent'' errors (\eg over-rotations in gates) behave very differently in random circuits, where they tend to get smeared out into effective depolarizing errors \cite{MagesanPRA12}, and in periodic circuits, where they can get amplified rapidly \cite{BlumeKohoutNC17}. More exotic faults like crosstalk and non-Markovianity are not yet well understood, but will almost certainly have strongly circuit-dependent effects. The circuits for algorithms and other applications tend to be highly and nontrivially structured. We do not yet fully understand how specific types of errors affect them! But we have no reason to believe that random and/or scrambling circuits will be a reliable proxy for algorithms. This is a strong argument for inventing and using other benchmarks -- ones that incorporate structured or periodic circuits, and ones that attempt to mimic real algorithm circuits. A diverse pool of benchmarks -- rather than a monoculture -- ensures robustness against unknown and unexpected effects.

Finally, quantum computing is still a very immature technology. There's a lot we don't know! Benchmarks will surely need to evolve. For example, it's not entirely clear how best to quantify ``success'' for a given circuit. Heavy outcome probability \cite{Aaronson16}, proposed for the quantum volume benchmark, is a promising idea -- but it's also one of the first ideas, and better success metrics may be invented. New error types may appear and become significant as hardware matures. New circuit types will almost surely be invented. Our goal in this paper is to encourage a proliferation of new, creative benchmarks from which (through use and argument) the best will emerge. Too many constraints inhibit creativity, but a total absence of structure inhibits communication. We hope the volumetric benchmark framework will hit a sweet spot, establishing a minimalist structure that enables all practitioners to measure and communicate performance, while allowing a lot of flexibility to define new benchmarks that capture many aspects of quantum computer performance. In the ``Examples'' (Sec.~\ref{sec:examples}), we show that many -- perhaps almost all -- extant benchmarks and test suites can be fitted into the volumetric framework, which makes us hopeful that it can encompass future benchmarks too.

\section{Framework for volumetric benchmarks}
\label{sec:framework}

A volumetric benchmark is:
\begin{enumerate}
\item A map from pairs of integers $(w,d)$ to ensembles of quantum circuits $\testsuite(w,d)$, all of width $w$ and depth $d$.
\item A rule detailing constraints on how a specified circuit may be \emph{compiled} into the native gates available on the target device.
\item A defined measure of ``success'' for each circuit, which takes as input a set of $N$ $w$-bit strings resulting from running that circuit on a device, and outputs either (a) a bit (0/1) indicating whether the device ``passed'' the test, or (b) a real number between 0 and 1 quantifying how \emph{well} it scored on the test.
\item A defined measure of overall success on an ensemble of circuits, which takes as input a set of success values (as defined in \#3, above) for all the circuits in an ensemble, and condenses them into a single representative number between 0 and 1.
\item \textbf{Optional} An \emph{experimental design} specifying how the circuits are to be run. The simplest experimental design is just a specification of $N$ (how many times each circuit should be run). More complicated designs may also include guidelines or requirements on what order to run the circuits in, such as whether the $N$ repetitions should be interleaved or divided in chunks to mitigate effects of drift \cite{VanEnkNJP13,Rudinger2018}, or perhaps even specifying an algorithm for choosing circuits adaptively based on preliminary results, etc.
\end{enumerate}

All of these properties should be specified and/or referenced clearly and unambiguously when a volumetric benchmark is reported. Some benchmarks (like randomized benchmarking) have evolved organically, and are performed in various ways \cite{Boone18}. For such benchmarks, repeatability and cross-platform comparison is only possible if the results are accompanied by a precise description of exactly what variation of the benchmark was performed (including, if at all possible, the order and timing of the circuits, which define the experimental design). If a benchmark has been precisely defined before (\eg in a published article), it is sufficient when reporting results to reference that standard and note any specific variations. When defining \emph{and} reporting the performance of a new benchmark, it is desirable to be as precise and specific as possible so that subsequent experiments can reproduce the same benchmark as closely as possible in order to enable comparison, or state precisely what was changed.

The width- and depth-dependent circuit ensemble $\testsuite(w,d)$ is the defining property of a benchmark. For a given circuit shape $(w,d)$, the test suite $\testsuite$ may contain:
\begin{enumerate}
\item A single circuit,
\item A list of $K(w,d)$ circuits, \emph{all} of which need to be run on the device,
\item A probability distribution over a large set of circuits, together with an integer $K(w,d)$ that specifies how many circuits should be drawn at random from the distribution and run on the device.
\end{enumerate}

Each circuit ensemble defines a \emph{test} that the device has to pass. The nature and specification of the circuit ensembles is up to the designer of the benchmark. However, the circuit ensembles are expected to share a common structure that ``scales'' naturally with $w$ and $d$. Increasing $w$ and/or $d$ is expected to make the test harder, or at least not easier. Several classes -- not necessarily exhaustive -- are discussed in Sec.~\ref{sec:classes}. Concrete examples of volumetric benchmarks from these classes are given in Sec.~\ref{sec:examples}.

The circuits defined by a given benchmark may need to be modified to run on real hardware. For instance, many platforms are unable to directly implement \CNOT gates. Instead, these gates must be synthesized from native two-qubit operations (such as \iSWAP or \MS gates) and single qubit operations. Conversely, automated circuit optimizers may be able to identify larger-scale structures in a VB's circuit ensemble that can be creatively recompiled to dramatically simplify their implementation. As an extreme example, each of the circuits used in randomized benchmarking composes to an identity operation. If the circuit optimizer were permitted to exploit this structure, the quantum computer could implement the benchmark without actually doing \emph{anything}. To avoid such situations, VBs need explicit rules constraining how the quantum circuits are to be implemented. We discuss this further in Sec.~\ref{sec:compilation}.

The other main ingredients of a benchmark define what it means to ``pass'' a test. First, what is required to succeed at performing a single circuit? Second, what counts as success for an ensemble of them? For single circuits, the simplest measure of success is ``Did running the circuit yield the correct/expected outcome more than $2/3$ of the time?'' (Note: $2/3$ is an arbitrary constant, but fairly conventional). However, this only makes sense for \emph{definite-outcome circuits} that have a unique ``correct'' outcome. Other criteria are discussed below in Sec.~\ref{sec:success}.

If a benchmark defines a single circuit for each shape, then ``passing'' just requires succeeding at that circuit. If the benchmark specifies a list of circuits, it may require \emph{all} circuits to be run successfully, or may require the \emph{average} success rate to be greater than $2/3$, or may require success on a fixed fraction of the circuits, \emph{etc}. Similar criteria apply to distributions over circuits, but in this case, average success rate is more compelling than (e.g.) criteria that require all circuits to succeed, because of reproducibility concerns when the circuits are drawn randomly and therefore unreproducibly. More complex criteria are also possible -- we do not intend to constrain the possible definitions. Potential success criteria (and their uses) are discussed below in Sec.~\ref{sec:success}.

Finally, a benchmark may or may not specify a particular experimental design. For example, randomized benchmarking is conventionally not associated with an experimental design -- the circuits can be run in any order, with the $N$ repetitions of each circuit interleaved or batched together as the experimentalist finds convenient. But in a benchmark designed to test stability over time (for example), a particular experimental design that specifies a particular order of circuits (\eg $N/2$ repetitions at one time, and $N/2$ at another time) might be important \cite{Rudinger2018}. We expect that most benchmarks will not require any special experimental design, but this allows for extensibility to address new tasks not envisioned at this time.


\section{Volumetric benchmarking circuits}
\label{sec:vbcircuits}
\begin{figure}[h!]
\centering
\includegraphics[width=0.85\linewidth]{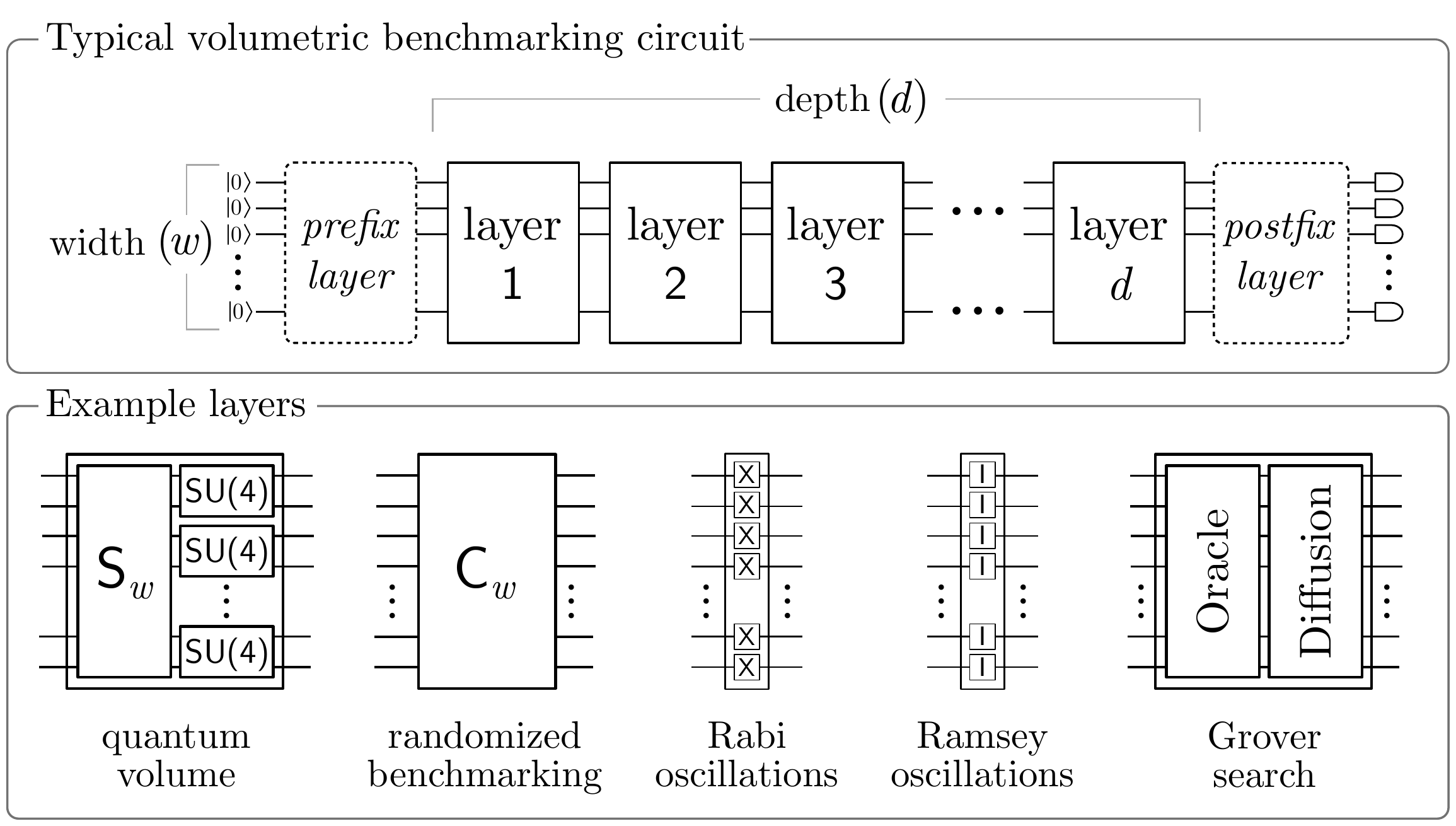}
\caption{(Top) The circuits used by volumetric benchmarks. Associated with each circuit is a width ($w$) equal to the number of qubits, and a depth ($d$) equal to the number of layers. Some circuits may also specify prefix and postfix operations. (Bottom) Examples of the layers used by various volumetric benchmarking circuits. Quantum volume layers are defined by a random permutation followed by pairwise random $\mathsf{SU(4)}$ elements. RB circuit layers are random $w$-qubit Clifford gates. Benchmarks based on Rabi oscillations can be composed of sequences of Pauli $X$ gates, while Ramsey oscillation benchmarks simply repeat a finite-duration identity gate on all qubits (with each qubit prepared and measured on the axis of the Bloch sphere). Benchmarks based on Grover search can be constructed using a layer composed of an oracle call and the Grover diffusion operator.}
\label{fig:depth}
\end{figure}

In order for a particular ensemble of quantum circuits to be useful as a volumetric benchmark, it must possess a well-defined notion of \emph{width} and \emph{depth}. Consequently, in this paper we \emph{only} consider circuits that can be specified using a sequence-of-layers structure, as illustrated in Fig.~\ref{fig:depth}. The width of such a circuit is simply the number of qubits required to implement the circuit, and the depth is the number of layers. Additional prefix and postfix layers may be included to prepare initial states and final measurements, such as the $\pi/2$ gates used with Ramsey benchmarks. These prefix/postfix layers are often qualitatively different from (and simpler than) the central layers, and are generally not counted towards the depth. So a circuit containing \emph{only} prefix and postfix layers would be said to have depth $d=0$.

The set of permissible central layers is essential to this definition of depth, and can vary dramatically between benchmarks. The following three examples illustrate the range of possibilities, and capture what we expect to be common use cases:
\begin{enumerate}
\item \textbf{Pseudocode circuits}:  Circuits can be specified at a very high level in terms of a set of layers that are actually subroutines.  Example subroutines include arbitrary Cliffords, ``controlled-$U$'' gates, QFTs, $n$-qubit Toffoli gates, Grover iterations, and adders.  This representation is roughly analogous to classical pseudocode -- it is relatively human-readable, and generally requires each layer to be extensively and creatively compiled to run on actual processors.  The quantum volume benchmark is specified at this level, as is standard randomized benchmarking (although its analysis often uses more concrete implementation details, like the average number of native gates per Clifford).
\item \textbf{Canonical-gate circuits}:  Circuits can be specified in terms of layers formed by parallel combination of ``canonical'' 1- and 2-qubit gates.  A typical example is arbitrary $SU(2)$ rotations on each qubit and CNOTs between each pair.  A physical device \emph{could} plausibly implement all these gates natively, but most real devices won't, and require compilation.  Each canonical gate can usually be compiled individually, via a standardized technique or lookup table.  Compiling a single canonical gate may produce a fairly complicated subcircuit (\eg \CNOT gates between widely separated qubits in a local architecture, or arbitrary rotations in an architecture with only discrete gates), but this complexity is usually controlled and predictable.
\item \textbf{Native-gate circuits}:  Circuits specified using layers formed by parallel combination of only the actual native gates available on a specific processor -- \eg specific 1-qubit gates, plus entangling gates between physically adjacent qubits -- are relatively concrete.  Compilation at this level is generally unnecessary, though there may be some freedom to \emph{schedule} those gates in time, by (1) inserting idle operations as desired, and (2) ``sliding'' gates forward and backward in time without exchanging the order of any non-idle gates. 
\end{enumerate}

The diversity of possible central layers highlights something important about volumetric benchmarks:  \textbf{performance on two different volumetric benchmarks is generally \emph{not} directly comparable}.  A 3-qubit processor might easily achieve $d=400$ on benchmarks derived from Rabi oscillation or direct RB circuits, but only $d=50$ on random-Clifford benchmarks, and top out at $d=10$ on benchmarks based on quantum volume circuits or Grover's algorithm.  Because the depth $d$ is determined from a high-level specification of the benchmarking task, all of these results are consistent. But even with two benchmarks whose layers are sampled from the same set, it usually doesn't make sense to compare performance directly.  For example, error can add up very differently in random circuits and periodic circuits, so it should not be surprising if a processor can achieve $d=1000$ on a random-circuit benchmark but only $d=200$ on a periodic benchmark \emph{or vice-versa}.  Certain specific cross-benchmark comparisons will make sense (\eg the last example would hint at coherent errors), but in general each benchmark will stand alone.

\subsection{Classes of circuits for volumetric benchmarks}
\label{sec:classes}

This paper presents a \emph{framework} for volumetric benchmarks, and is not intended to introduce (or advocate for) any particular benchmarks. We hope to see a wide variety of benchmarks proposed, each motivated by a different aspect of ``performance'' that the benchmark seeks to capture. The possibility of basing benchmarks on creative and novel kinds of circuits is essential for that vision. But we also want to illustrate the framework's potential with some examples, so in this section we give a non-exhaustive list of circuit classes that could (and in some cases already do) form benchmarks. 

\subsubsection{Random circuits}
\label{sec:classes_randomcircuits}

Random circuits may be the most obvious choice, and many ensembles of random circuits are known. Generally, a random circuit class is defined by (1) specifying an ensemble of subroutine circuits, and then (2) building a large circuit of width $w$ and depth $d$ by drawing subroutines at random and combining them in series, parallel, or both. Some higher-level structure is usually imposed as well.  For example, standard (Clifford) randomized benchmarking circuits \cite{MagesanPRL11} are mostly composed of randomly chosen $w$-qubit Clifford subroutines combined in series, but with the high-level constraint that the last Clifford must invert the previous $d-1$ Cliffords.

Clifford RB is not the only random-circuit volumetric benchmark. Direct RB \cite{Proctor18} replaces the random Clifford subroutines with $d$ depth-1 layers that are themselves formed by parallel combination of native 1- and 2-qubit gates on $w$ qubits. Here, the higher-level structure takes the form of a prefix and a postfix. The prefix prepares a random stabilizer state and the postfix transforms to the computational basis before measurement. (Note that ``depth'' $d$ has a different meaning in each of these, as discussed above). The simultaneous RB protocol \cite{GambettaPRL12} defines a benchmark composed of circuits that combine 1- or 2-qubit Clifford (or direct) RB subroutines in parallel across $w$ qubits. There are several other variations and spin-offs of RB, \emph{\eg} cycle benchmarking \cite{Erhard19}, that could also be adapted as volumetric benchmarks.

Conventionally, RB is used as a QCVV protocol to measure an intrinsic error rate that can be extracted through particular data analysis. Here, we ignore that application, and propose repurposing the RB circuits as a volumetric benchmark. In this use, the results are not analyzed in the conventional way; we propose a standardized way to report volumetric benchmarking in Sec.~\ref{sec:reporting_results}.

\emph{Scrambling} circuits, which are used in quantum volume \cite{Cross18,SekinoJHEP08}, are another class of random circuits.  They can, in principle, explore \emph{every} corner of the system's Hilbert space, while RB circuits typically only explore a discrete subset that spans the state space.  Scrambling circuits can be constructed in various ways. The construction given by IBM for quantum volume involves a high-level structure: the circuits are composed of $d$ alternating sequential layers, each of which is either a randomly chosen subroutine that permutes the $w$ qubits or a randomly chosen subroutine that applies local 2-qubit unitaries from $SU(4)$.  Another class has been proposed by Google \cite{BoixoNP18,NeillScience18}.

\subsubsection{Periodic circuits}
\label{sec:classes_periodiccircuits}

The oldest qubit test suites -- Rabi and Ramsey sequences -- are periodic rather than random.  More recent test protocols that define suites of periodic circuits include robust phase estimation (RPE) \cite{KimmelPRA15} and gate set tomography (GST) \cite{BlumeKohoutNC17}.  Periodic circuits repeat a single subroutine $O(d)$ times.  They stress-test different aspects of a processor's performance from the ones emphasized by random circuits -- coherent errors can be amplified by periodic circuits, whereas in random circuits they tend to get twirled away or smeared out.  Even the simplest periodic circuit, a long sequence of ``idle'' operations, each of which is intended to do nothing for a single clock cycle, is useful for identifying calibration errors and decoherence.  It's easy to envision a processor that excels at random circuits but fails at periodic ones, or vice versa.  Doing equally well on \emph{both} kinds of benchmark is a hallmark of reliability.

Periodic circuits are also useful because they reflect the structure of many quantum algorithms.  Grover's algorithm is an extreme example; two subroutines are performed alternately $O(\sqrt{N})$ times to search a set of $N$ possible solutions.  Phase estimation, a very common subroutine \cite{ClevePRSA98}, relies on repeated applications of a unitary to amplify its eigenvalues.

\subsubsection{Application circuits}
\label{sec:classes_applicationcircuits}
Not all algorithms are periodic or random.  Furthermore, generic periodic circuits (Rabi/Ramsey sequences or GST circuits) don't reflect the complexity of algorithm circuits.  So a third essential class of circuits is \emph{application circuits} -- circuits that behave like the ones that are expected to be run in real-world usage.  Currently, applications for circuit-model quantum computers fall into three main classes: ``digital'' quantum algorithms that are intended to be compiled into discrete gate sets to run on error-corrected qubits; ``analog'' quantum algorithms designed to run on physical qubits that admit continuous families of native gates (\eg $Z$ rotations by $\theta$) \cite{PeruzzoNC14}; and quantum error correction.

In certain cases, it may be possible to define a volumetric benchmark by simply choosing actual instances of a given algorithm that correspond to each $(w,d)$ pair.  However, most algorithms don't scale smoothly to every depth and width.  For example, full Grover search on $w$ qubits requires depth $O(2^{w/2})$.  But the circuits in a benchmark needn't actually perform an algorithm.  Rather, they should be \emph{representative} of the circuits that perform that algorithm (or other application).  For example, a Grover-inspired benchmark might include circuits that perform $d$ iterations of the Grover subroutine -- for any integer $d$ -- in between ``prefix'' and ``postfix'' subroutines designed to simplify the circuits' output distribution.  More generally, a wide variety of representative circuits can be constructed by identifying key subroutines for larger applications (\eg phase estimation or syndrome extraction) that scale more smoothly with $d$ and $w$ than the full-scale application.

Eventually, application circuits will constitute the most important and useful class of test suites.  However, identifying the best ways to construct test suites based on applications will require significant additional research.  As we said above, our goal in this paper is to construct a useful container for benchmarks, not the benchmarks themselves. We anticipate that the volumetric framework and the observations above will help to guide the construction of application-specific benchmarks.

\section{Rules for compiling volumetric benchmarking circuits}
\label{sec:compilation}
Quantum circuits are an expressive and versatile language for specifying a wide range of quantum programs and subroutines. But a quantum circuit alone is almost never sufficient to specify exactly how a program will be implemented on real hardware. Most quantum circuits require some degree of compilation into explicit machine instructions. The flexibility granted to the compiler will necessarily depend on the available hardware resources, but also on the purpose of the benchmark. Benchmarks intended to capture higher-level performance characteristics (\eg algorithmic benchmarks) will permit more flexibility in their implementation than benchmarks intended to reflect specific, low-level hardware details (\eg Ramsey circuits). 

Most benchmarks permit \emph{compiling up} -- increasing the complexity of the circuit to accommodate hardware constraints, such as limited connectivity or a small gate set.  But many benchmarks restrict \emph{compiling down} -- utilizing conventional computation to simplify the circuit. Many (though not all) compiler restrictions can be represented by inserting explicit \emph{barriers} in the quantum circuit specification, as in Fig.~\ref{fig:barriers}. These permit the compiler to optimize the circuit locally, between between each pair of adjacent barriers, but forbid it from using information about other parts of the circuit. These barriers might be chosen to reflect the explicit layer structure of the circuit ensemble, or they may not. 

\begin{figure}[h!]
    \includegraphics[width=\columnwidth]{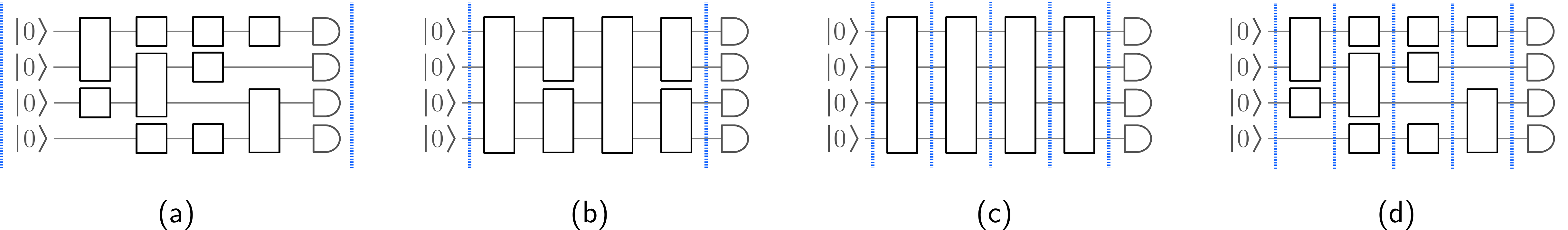}
    \caption{Specification of compiler constraints can often be aided by including barriers (blue lines above) in quantum circuit definitions. (a) Google's random circuit benchmark allows for the entire quantum circuit to be recompiled as desired, up to and including replacement of the quantum computation with a classical simulation. (b) IBM's quantum volume circuits are specified using a particular family of quantum circuits, but the described unitary operation may be recompiled freely. (c) Randomized benchmarking specifies a sequence of unitary operations that may be individually compiled into native gates, but not jointly compiled. (d) Direct randomized benchmarking specifies circuit layers composed of native gates, and the only compilation permitted is to accommodate hardware constraints on parallelism. Alternatively, one may choose to explicitly forbid serialization within the layers (\eg in crosstalk benchmarking).}
    \label{fig:barriers}
\end{figure}

The following examples illustrate the diversity of compilation restrictions in existing benchmarks, ordered by increasing rigidity of the compiler:
\begin{enumerate}
    \item \textbf{Random circuit sampling}~\cite{BoixoNP18,NeillScience18}: For benchmarking purposes, random quantum circuits can be interpreted as specifying distributions over bit strings, defining a benchmark whose goal is to sample from these distributions as accurately as possible. This benchmark admits tremendous flexibility in compilation, up to and including replacing the quantum computation with a conventional simulation. This flexibility allows the benchmark to be performed on both conventional and quantum hardware. In fact, Google was able to make its claim of quantum supremacy~\cite{AruteNature19} by running a particular instance of this benchmark on quantum hardware that could not be run on a conventional supercomputer in a reasonable amount of time. See Fig.~\ref{fig:barriers}(a).
    \item \textbf{Quantum volume}~\cite{Cross18}:  Quantum volume circuits are used as a convenient mechanism for specifying particular unitary operations. The goal of the benchmark is to implement these unitaries accurately, but the ultimate machine instructions need not resemble the specification at all. In fact, this benchmark encourages compilations that merely \emph{approximate} the target unitary, as long as the approximation error is offset by a concomitant reduction in hardware errors. One should, however, make a good faith attempt to implement the unitary, rather than simply mocking up the correct output distribution based on the results of a classical simulation. See Fig.~\ref{fig:barriers}(b).
    \item \textbf{Randomized benchmarking}~\cite{MagesanPRL11,MagesanPRA12} - Randomized benchmarking circuits are typically specified as a series of multi-qubit Clifford gates that compose to the identity. Each individual Clifford operation must generally be compiled from native gates, but compilation between layers is forbidden, lest the entire circuit be compiled down to nothing. See Fig.~\ref{fig:barriers}(c).
    \item \textbf{Direct randomized benchmarking}~\cite{Proctor18} - Direct randomized benchmarking circuits use layers of native Clifford-generators that compose to a random Pauli gate. As above, no compilation is permitted between layers or the circuit would become trivial. Within each layer, no compilation is required, except for the possible inclusion of idle operations that may be inserted as necessary to satisfy hardware restrictions on parallelism. See Fig.~\ref{fig:barriers}(d).
    \item \textbf{Crosstalk detection protocols}~\cite{GambettaPRL12} - Crosstalk detection benchmarks are often constructed so that specific gates occur simultaneously in the circuit. No compilation should be required. In particular, the compiler should \emph{not} be permitted to serialize the gate operations. See Fig.~\ref{fig:barriers}(d).
\end{enumerate}

As evident from these examples, compilation constraints can vary dramatically between benchmarks. Benchmarks that permit extensive compilation are easily adapted to different hardware platforms, and their results are reflective both of core quantum performance and of compiler performance. On the other hand, benchmarks that restrict compilation provide more direct information about the low-level performance of the quantum hardware. 

However, low-level hardware implementation details can pose a problem when explicitly constraining the compilation freedom. Some modern devices implement all physical $Z$ rotations ``virtually'', essentially by recording a change of reference frame for the qubit experiencing the $Z$-rotation, and then propagating it through subsequent operations \cite{IBMZGate}.  This is clever, beneficial, and entire legitimate -- but a computer that implements it literally \emph{cannot} implement a $Z$ gate without compilation that bridges multiple operations, since $Z$ gates are explicitly implemented by altering all subsequent $X$ or $Y$ rotations. Such compilation is generally benign, unless, for example, the benchmark circuits consist primarily of many repeated $Z$ gates. Similar complications arise in superconducting platforms that use pulse distortion to correct for gates that appear earlier or later in the circuit~\cite{Rol2020-dp}. 

The description of a volumetric benchmark should be clear about how much implementation freedom -- \eg compilation, scheduling, etc -- the benchmark permits.  This freedom should be consistent with (a) the specific processor being tested, and (b) the gate set used to specify circuits. If those gates \emph{require} a particular form of compilation for implementation on that processor, then the benchmark can only be performed if that form is allowed.  Some benchmarks may not be compatible with some processors.

\section{Success criteria for individual circuits}
\label{sec:success}

Passing a test comes down to succeeding at individual circuits.  The most obvious definition of success at circuit $C$ is: \textbf{Running $C$ on an as-built processor \emph{always} yields the same result that it would on a perfect processor with no errors.}  But this definition is too strict -- some amount of failure has to be tolerated in order to evaluate today's processors -- and it only applies to \emph{definite-outcome circuits} that would ideally produce a unique outcome.  Most circuits are intended to generate random outputs sampled from a specific but nontrivial probability distribution.

Several success criteria are available, and we expect more to be invented.  It's critical that when benchmarks are defined and used, the success criterion that was used is (1) specified precisely, and (2) motivated and/or explained.  Since there are as yet no hard and fast rules for what makes something a ``useful'' or valid success criterion, there is potential for abuse -- someone could in principle introduce benchmarks based on weak or meaningless success criteria, and use them for ``benchmarketing'' (inflating performance figures in a misleading fashion).  The defense against this abuse is to demand a clear motivating description of \emph{why} a new criterion really does indicate ``success'', and how it compares to other, better studied, criteria.

Here are some candidate criteria:
\begin{itemize}
\item The probability of the correct outcome of circuit $C$ is greater than a specific threshold (\eg $2/3$) with high (\eg $95\%$) statistical confidence.  Note that this only applies to definite-outcome circuits, and see discussion below.
\item The \emph{heavy output probability} \cite{Aaronson16} of circuit $C$ is greater than a specific threshold (\eg $2/3$) with high statistical confidence.  This is the criterion used for quantum volume \cite{Cross18}.  Note that this applies to some indefinite-outcome circuits, but not \emph{uniform-outcome circuits} (where all outcomes are equally probable), and also that it requires the ability to compute all outcome probabilities in advance.
\item The distribution of the outcomes of circuit $C$ is within a specific distance of the ideal distribution (according to some specified metric such as total variational distance) with high statistical confidence.  This criterion can apply to all circuits (definite-, indefinite-, and uniform-outcome), but requires computing outcome probabilities in advance. Also, the amount of data required to achieve statistical significance increases with the entropy of the distribution.
\item The \emph{coarse-grained} outcome distribution is close to ideal (precisely as stated in the previous criterion) with respect to some specific coarse-graining.  See discussion below -- this is a powerful class of criteria that includes heavy output probability as a special case.
\item The probability of an output within a specified Hamming distance (\eg 2) of the correct outcome is greater than a certain threshold (\eg $2/3$).  This definition captures the idea that in a many-qubit processor, some results are more wrong than others. If only few bits ever get flipped, this suggests that the processor may be suitable for quantum error correction.
\item The observed outcome frequencies of circuit $C$ are \emph{consistent} with the predicted probabilities, meaning that the empirical cross-entropy between them is below a specified threshold.  This interesting quantifier of success is at the heart of Google's proposal for demonstrating quantum supremacy \cite{NeillScience18,BoixoNP18}.
\end{itemize}

A rather long paper could be written entirely about success criteria.  We do not have space to discuss this topic in that level of detail!  Our goal here is to state, and demonstrate, the simple idea that \emph{there are many useful ways to define ``success'' at running a circuit}.  Most of the definitions given here have appeared in testing and characterization protocols already.

Two points are worth touching on briefly, though.  First, as noted in the first criterion (probability of the correct outcome), even the simplest criteria have some subtleties.  The precise \emph{probability} of a given outcome is not accessible through experiment.  Flipping a coin 1000 times and observing 689 heads doesn't mean that $Pr(\mathrm{heads}) = 0.689$.  We can only make statements such as:  given the \emph{empirical frequency} of $0.689$, $Pr(\mathrm{heads})$ is greater than $2/3$ with reasonably high confidence, and greater than $1/2$ with very high confidence.  It's also not clear where to put the threshold.  $2/3$ is common in computer science, but this is explicitly arbitrary.  The best-motivated threshold may be problem-dependent.

Second, we note that \emph{coarse-graining} of the output distribution is a powerful and general technique.  This means taking the $2^w$ possible outcomes of a circuit on an $w$-qubit processor and dividing them into $k \ll 2^w$ bins according to some rule.  Now we can discuss, report, and analyze the empirical and predicted distributions over just $k$ outcomes.  If $k$ is kept constant as $w$ increases, then many problems with comparing those distributions become manageable.

There are many ways to coarse-grain.  Heavy output probability, which is used in quantum volume \cite{Cross18} and also in other quantum supremacy benchmarks \cite{Aaronson16,BoulandNP19}, is one example.  By computing all the outcome probabilities in advance, the outcomes are divided into just two bins:  those with probability greater than the median, and those with probability less than the median.  Now, evaluating success is simply a matter of comparing two \emph{2-outcome} distributions (the empirical one and the predicted one).  This is a neat and useful idea -- but it's not unique.  Many other useful coarse-grainings exist, and should be considered.  One is local marginals (obtained by marginalizing over all bits except one).  Or, for an algorithm that is supposed to produce an outcome that's uniformly random over a relatively small subset of strings, the natural coarse-graining is into (a) all the strings that \emph{should} occur, and (b) the rest.

\section{Success critera for circuit ensembles}
\label{sec:ensembles}

A volumetric benchmark defines a family of tests -- one for each circuit shape $(w,d)$ -- that a processor must pass.  A test can take various forms.  The simplest test is ``Demonstrate the ability to run a specific circuit $C(w,d)$ and get an acceptable answer most of the time.''  But there are other kinds of tests.  So a volumetric benchmark defines, for each circuit size $(w,d)$, an \emph{ensemble} of circuits that we denote $\testsuite(w,d)$.  The term ``ensemble'' should be interpreted broadly, encompassing at least three distinct kinds of test:
\begin{enumerate}
\item A single circuit.
\item A finite list of circuits, some or all of which must be ``passed''.
\item A distribution over a [possibly infinite] set of circuits, from which a sample is to be drawn.
\end{enumerate}
Passing a single-circuit test just requires demonstrating the ability to run that circuit successfully.  We discuss how to evaluate this in the next section below.  Benchmarks of this form will be relatively rare, because we usually want to know whether a processor can run many circuits of a particular form.  
Passing a list-of-circuits test can mean at least three distinct things:
\begin{enumerate}
\item Running all circuits in the list successfully.
\item Running at least a certain number (\eg $90\%$) of the circuits successfully.
\item Achieving an \emph{average} quantitative success -- defined by the average of some success metric (see next section) over all the circuits in the list -- above a specified threshold.
\end{enumerate}
We see the first criterion -- success at \emph{all} circuits -- as the most clearly useful.  While it is clearly the most challenging of the alternatives, it is by no means unreasonable.  All the circuits in $\testsuite(w,d)$ are of the same size, and it's reasonable to expect a quantum processor that can run some or most circuits of a given size to run \emph{all} circuits of that size.  Moreover, benchmarks that require success at all circuits provide strong guarantees of performance.  No user wants to find out (later) that the circuit they need to run is one of the few that their processor \emph{can't} run!

\section{Examples of volumetric benchmarks}
\label{sec:examples}

The motivation for the VB framework presented here is simple:  it gathers up and generalizes many existing techniques that have been proposed and used to evaluate the performance of quantum processors.  Every aspect of the framework discussed in the previous sections was inspired directly by some specific known protocol (although to the best of our knowledge the visualization and reporting paradigm illustrated in the next section is novel).  So we now give some examples of how existing protocols might be adapted into VBs.

We emphasize that the VB based on an existing protocol is not necessarily identical to the protocol itself, because the \emph{data analysis} is often different.  QCVV protocols generally define not just a  circuit family, but also a specific analysis routines that extract intrinsic properties of a device. For example, simultaneous RB data is usually analyzed via a nontrivial procedure to detect crosstalk, and gate set tomography data is usually fit to a complex Markovian gate set model. So when we co-opt a protocol's circuit definitions to use as a VB, the associated analyses are \emph{not} part of the VB framework. A VB based on GST might inherit the same circuit families used for GST, but analyze the data in a totally different (and much simpler) way that makes no attempt to infer a gate set.  Furthermore, it may be useful to generalize the circuit families defined by a given QCVV protocol. For instance, the standard GST analysis makes use only of circuits with depths equal to powers of two. When GST circuits are adapted for use as a volumetric benchmark this restriction can be easily relaxed to construct circuits of nearly any depth. 

As we observed above, circuit classes are the heart of a VB.  Changing the circuit class structure (\eg from random circuits to periodic ones) creates an entirely new and incomparable VB, whereas other changes (\eg to the success metric) may have little effect.  So we organize our list of examples by circuit type.  This is not intended to be exhaustive; there are undoubtedly some good examples that we fail to mention here.  For each example, we list all the specific properties and choices that define a VB, as well as a brief description of the device property that is measured by the benchmark.

\subsection{Random circuits}
\label{sec:ex_randomcircuits}
The examples given here, based on various types of random circuits, require almost no adaptation to fit in the VB framework.  Data from randomized benchmarking, scrambling (quantum volume) circuits, and random quantum supremacy circuits are usually analyzed in simple ways that are consistent with the VB framework.  Although RB data is often used to estimate a decay rate -- which goes outside the analysis we suggest for a VB -- this simple estimation of a parameter is very closely correlated with the VB analysis, which would simply identify the size/shape of quantum circuits where the overall failure probability drops below a pre-established threshold.

\noindent\rule{\columnwidth}{.5pt}
\noindent \textbf{Protocol:} \emph{Clifford randomized benchmarking} \cite{MagesanPRL11,MagesanPRA12}. 
\\ \noindent \textbf{Circuit family $\testsuite(w,d)$:} Sequences of arbitrary $w$-qubit Clifford subroutines, postfixed by a single fully-constrained inversion Clifford.
\\ \noindent \textbf{Measure over circuits:} Uniformly random over the first $d-1$ Cliffords.
\\ \noindent \textbf{Compilation rule:} Individual Clifford operations may be compiled without restriction, but no compilation across Cliffords is permitted (see Fig.~\ref{fig:barriers}(c)).  Users are encouraged to report the details of their Clifford compilation. Standard compilations will result in circuits with a physical depth $=O(dw/\log w)$ for $w$ qubits, or higher if the device connectivity is not all-to-all.
\\ \noindent \textbf{Per-circuit success metric:}  Probability of unique correct outcome.
\\ \noindent \textbf{Family-wise success metric:}  Uniform average over all sampled circuits.
\\ \noindent \textbf{Property probed:}  Average error of Clifford circuits.

\noindent\rule{\columnwidth}{.5pt}
\noindent \textbf{Protocol:} \emph{Direct randomized benchmarking} \cite{Proctor18}.
\\ \noindent \textbf{Circuit family $\testsuite(w,d)$:} Sequences of arbitrary $w$-qubit depth-1 layers chosen from a set that generates the $w$-qubit Clifford group, prefixed and postfixed by random-stabilizer-initialization and inversion Clifford subroutines.
\\ \noindent \textbf{Measure over circuits:} Random over sequences of layers, i.i.d. with respect to each layer, but with adjustable constraints and probability distributions on layers (see citation for details).
\\ \noindent \textbf{Compilation rule:} Each layer may be compiled as necessary, but compilation across layers is forbidden (see Fig.~\ref{fig:barriers}(d)).
\\ \noindent \textbf{Per-circuit success metric:}  Probability of unique correct outcome.
\\ \noindent \textbf{Family-wise success metric:}  Uniform average over all sampled circuits.
\\ \noindent \textbf{Property probed:}  Average error of randomized native-gate circuits.

\noindent\rule{\columnwidth}{.5pt}
\noindent \textbf{Protocol:} \emph{Simultaneous randomized benchmarking} \cite{GambettaPRL12}. 
\\ \noindent \textbf{Circuit family $\testsuite(w,d)$:} Sequences of arbitrary $w$-parallel $1$-qubit Clifford subroutines, each postfixed by a single fully-constrained inversion Clifford.
\\ \noindent \textbf{Measure over circuits:} Uniformly random over the first $d-1$ Cliffords, and independent for each qubit.
\\ \noindent \textbf{Compilation rule:} Parallel Clifford operations may be compiled without restriction, but compilation across non-parallel Cliffords is forbidden. Users of SRB benchmarks should explicitly indicate if extra idle operations are inserted to avoid parallelization of active operations.  
\\ \noindent \textbf{Per-circuit success metric:}  Probability of unique correct outcome.
\\ \noindent \textbf{Family-wise success metric:}  Uniform average over all sampled circuits.
\\ \noindent \textbf{Property probed:}  Average error of non-entangling (local) circuits.

\noindent\rule{\columnwidth}{.5pt}
\noindent \textbf{Protocol:} \emph{Quantum volume benchmark} \cite{Cross18}. 
\\ \noindent \textbf{Circuit family $\testsuite(w,d)$:}  $w$-qubit scrambling circuits that alternate layers of arbitrary permutations of the qubits with nearest-neighbor 2-qubit arbitrary $SU(4)$ subroutines.
\\ \noindent \textbf{Measure over circuits:} Uniformly random over the permutation group $\pi(w)$ and $SU(4)^{\otimes w/2}$.
\\ \noindent \textbf{Compilation rule:} Each circuit specified by this benchmark should be interpreted as defining a target unitary operation. Users are permitted to recompile this unitary operation without restriction (see Fig.~\ref{fig:barriers}(b)).
\\ \noindent \textbf{Per-circuit success metric:}  Heavy outcome probability $>2/3$.
\\ \noindent \textbf{Family-wise success metric:}  Average of single-circuit criterion.
\\ \noindent \textbf{Property probed:}  Ability to run ``square'' scrambling circuits (prepare arbitrary/random $w$-qubit states).

\noindent\rule{\columnwidth}{.5pt}
\noindent \textbf{Protocol:} \emph{Google cross-entropy benchmark} \cite{BoixoNP18,NeillScience18}. 
\\ \noindent \textbf{Circuit family $\testsuite(w,d)$:} $w$-qubit scrambling circuits that alternate layers of $1$-qubit gates and $2$-qubit gates.
\\ \noindent \textbf{Measure over circuits:} complicated; see reference.
\\ \noindent \textbf{Compilation rule:} Each circuit specified by this benchmark should be interpreted as defining a probability distribution over output bitstrings. Users are permitted to recompile the circuit as desired, up to and including replacement of the quantum computation with a classical approximate weak simulation (see Fig.~\ref{fig:barriers}(b)). 
\\ \noindent \textbf{Per-circuit success metric:}  Cross-entropy between predicted probabilities and empirical frequencies.
\\ \noindent \textbf{Family-wise success metric:}  Average of single-circuit criterion.
\\ \noindent \textbf{Property probed:}  Ability to run classically hard-to-simulate scrambling circuits. 

\noindent\rule{\columnwidth}{.5pt}

\subsection{Periodic circuits}
\label{sec:ex_periodiccircuits}

The VB examples given here are also well-established and canonical.  However, unlike the random-circuit examples above, they are \emph{not} usually viewed as ``benchmarks''.  More commonly, the data produced by these protocols -- Rabi/Ramsey oscillations, sequences of idles, RPE, and GST -- is analyzed in complicated ways to estimate parameters such as gate rotation angles, frequencies, or process matrices.  Here, we suggest success metrics that could be used to cram these well-known protocols into the VB framework, and identify the property that they probe when so used.

\noindent\rule{\columnwidth}{.5pt}
\noindent \textbf{Protocol:} \emph{Rabi oscillations}. 
\\ \noindent \textbf{Circuit family $\testsuite(w,d)$:} $d$ consecutive repetitions of a single layer of local $1$-qubit gates.
\\ \noindent \textbf{Measure over circuits:} Unique circuit.
\\ \noindent \textbf{Compilation rule:} No compilation permitted.
\\ \noindent \textbf{Per-circuit success metric:}  Total variational distance between predicted and observed local outcome probabilities $<$ threshold.
\\ \noindent \textbf{Family-wise success metric:}  N/A
\\ \noindent \textbf{Property probed:}  Over/under-rotation angle and decoherence/error rate of a single quantum gate/layer.

\noindent\rule{\columnwidth}{.5pt}
\noindent \textbf{Protocol:} \emph{Ramsey oscillations}. 
\\ \noindent \textbf{Circuit family $\testsuite(w,d)$:} $d$ consecutive repetitions of a single layer of idles or local $1$-qubit Z-rotations, pre- and post-fixed by depth-1 state-preparation and measurement layers.
\\ \noindent \textbf{Measure over circuits:} Unique circuit.
\\ \noindent \textbf{Compilation rule:} No compilation permitted.
\\ \noindent \textbf{Per-circuit success metric:} Total variational distance between predicted and observed local outcome probabilities $<$ threshold. 
\\ \noindent \textbf{Family-wise success metric:} N/A 
\\ \noindent \textbf{Property probed:}  Over/under-rotation angle and dephasing/error rate of a single idle or rotation gate.

\noindent\rule{\columnwidth}{.5pt}
\noindent \textbf{Protocol:} \emph{Idle tomography} \cite{BlumeKohout19}. 
\\ \noindent \textbf{Circuit family $\testsuite(w,d)$:} $d$ consecutive repetitions of a $w$-qubit idle, pre- and post-fixed by a specific ensemble of $O(\log w)$ depth-1 state-preparation and measurement layers.
\\ \noindent \textbf{Measure over circuits:} Explicit list of $K = O(\log w)$ circuits; all must be run.
\\ \noindent \textbf{Compilation rule:} No compilation permitted.
\\ \noindent \textbf{Per-circuit success metric:}  Total variational distance between predicted and observed 1- and 2-local marginal outcome probabilities $<$ threshold.
\\ \noindent \textbf{Family-wise success metric:}  All individual circuits must succeed.
\\ \noindent \textbf{Property probed:}  Rate of worst errors in an $n$-qubit idle operation.

\noindent\rule{\columnwidth}{.5pt}
\noindent \textbf{Protocol:} \emph{Robust phase estimation} \cite{KimmelPRA15}. 
\\ \noindent \textbf{Circuit family $\testsuite(w,d)$:} Specific Rabi/Ramsey-type circuits (see reference), parallelized over $w$ qubits.
\\ \noindent \textbf{Measure over circuits:} Explicit list of circuits; all must be run at each chosen depth.
\\ \noindent \textbf{Compilation rule:} No compilation permitted. 
\\ \noindent \textbf{Per-circuit success metric:}  Total variational distance between predicted and observed 1- and 2-local marginal outcome probabilities $<$ threshold.
\\ \noindent \textbf{Family-wise success metric:}  All individual circuits must succeed.
\\ \noindent \textbf{Property probed:}  Over/under-rotation angle and decoherence/error rate of one or more quantum gates.

\noindent\rule{\columnwidth}{.5pt}
\noindent \textbf{Protocol:} \emph{Long-sequence gate set tomography} \cite{BlumeKohoutNC17}. 
\\ \noindent \textbf{Circuit family $\testsuite(w,d)$:} Specific Rabi/Ramsey-type circuits involving repetitions of specific depth-$O(1)$ ``germs'' (see reference), parallelized over $w$ qubits.
\\ \noindent \textbf{Measure over circuits:} Explicit list of circuits; all must be run at each chosen depth.
\\ \noindent \textbf{Compilation rule:} No compilation permitted.
\\ \noindent \textbf{Per-circuit success metric:}  Various possible.  Simplest is total variational distance between predicted and observed 1- and 2-local marginal outcome probabilities $<$ threshold.
\\ \noindent \textbf{Family-wise success metric:}  All individual circuits must succeed.
\\ \noindent \textbf{Property probed:}  Rate of worst errors in a set of 1- or 2-qubit quantum gates.

\noindent\rule{\columnwidth}{.5pt}

\subsection{Application circuits}
\label{sec:ex_applicationcircuits}

The example VBs listed in this section are speculative.  We are not aware of any experiments or explicit proposals that could be evaluated as a VB in current form.  However, for the two circuit families we suggest -- iterations of a Grover step, and steps of a Trotterized Hamiltonian simulation -- we believe it is obvious that a VB \emph{could} be built around them.  Exactly how to do so is not obvious, though.  Many variations are possible, and we do not attempt to nail down the specifics here.  Establishing the best way to construct circuits for these VBs, and identifying other algorithms/applications that can form the basis of VBs, is a promising area for future work.  (Note:  VBs based on syndrome extraction circuits for quantum error correction are also good candidates for ``application-focused'' benchmarks, but identifying a sensible circuit family is nontrivial for several reasons.  For example, in contrast to all other examples, increasing the number of physical qubits $w$ is likely to \emph{increase} the success probability if it increases the code distance.) 

\noindent\rule{\columnwidth}{.5pt}
\noindent \textbf{Protocol:} \emph{Grover iterations} \cite{GroverPRL97}. 
\\ \noindent \textbf{Circuit family $\testsuite(w,d)$:} $d$ iterations of a single $w$-qubit Grover step, alternating oracle marking and reflection steps.  Oracle can be defined by user; simplest choice is to just mark $\ket{0}$ or a randomly chosen fixed basis state. The physical depth of the circuits will depend strongly on architecture, and even more strongly on the oracle chosen if it is nontrivial.
\\ \noindent \textbf{Measure over circuits:} Unique circuit unless the oracle is allowed to vary; for the simple oracle suggested above, the marked state could be uniformly random.
\\ \noindent \textbf{Compilation rule:} The specified unitary operation may be compiled freely (see Fig.~\ref{fig:barriers}(b)). 
\\ \noindent \textbf{Per-circuit success metric:}  Heavy outcome probability, cross-entropy, or other metric suitable for arbitrary non-definite outcome distributions.
\\ \noindent \textbf{Family-wise success metric:}  Average of single-circuit criterion (if applicable)
\\ \noindent \textbf{Property probed:}  Ability to run Grover's algorithm without errors.

\noindent\rule{\columnwidth}{.5pt}
\noindent \textbf{Protocol:} \emph{Trotterized Hamiltonian simulation} \cite{LloydScience96}. 
\\ \noindent \textbf{Circuit family $\testsuite(w,d)$:} $d$ iterations of a single Trotter step for simulating a $w$-qubit Hamiltonian comprising a sum of local terms. The physical depth of the circuits depends strongly on architecture and even more strongly on the Hamiltonian chosen.
\\ \noindent \textbf{Measure over circuits:}  Various choices:  both input state and Hamiltonian could be varied either randomly over some measure or systematically.  Alternatively, a single physically-motivated Hamiltonian and input state could be used (with $w$ controlling fineness of discretization).
\\ \noindent \textbf{Compilation rule:} The specified unitary operation may be compiled freely (see Fig.~\ref{fig:barriers}(b)).
\\ \noindent \textbf{Per-circuit success metric:}  Heavy outcome probability, cross-entropy, or other metric suitable for arbitrary non-definite outcome distributions.  Alternatively, error in empirically estimated expectation value of an observable of physical interest must be $<$ threshold.
\\ \noindent \textbf{Family-wise success metric:}  Various choices if $\testsuite(w,d)$ is non-unique; simplest choice to to require all circuits to succeed.
\\ \noindent \textbf{Property probed:}  Duration over which Hamiltonian time evolution can be simulated accurately.

\noindent\rule{\columnwidth}{.5pt}

\section{Reporting and interpreting VB results}
\label{sec:reporting_results}

\noindent\textbf{\emph{Note: The figures in this section are for illustration purposes only.  They do not represent real data; the ``data'' presented in these figures do \underline{not} come from real devices or rigorous simulations!}}

In this section, we propose and illustrate a consistent, informative visual representation of data from volumetric benchmarking experiments.  It allows a single figure to provide a comprehensive answer to the nontrivial question ``What quantum circuits of a specific type can (and can't) this device run successfully?''  The value of a VB, as an alternative to single-number metrics like quantum volume, is the richness of the answer it provides to this question.  This richness is, in general, better suited to visual display. A useful volumetric benchmarking plot should allow a reader to quickly understand the gross performance of one or more quantum devices under some particular metric of success. At a minimum, it should address the following questions:
\begin{enumerate}
	\item What volumetric benchmark was used?
	\item For which width/depth pairs were experiments successful?
\end{enumerate}

We display benchmark results by arranging circuit shapes (width/depth pairs) in a grid, with circuit depth increasing along the abscissa (x-axis) and circuit width increasing along the ordinate (y-axis).  We specify the circuit family and success metric in the figure title and caption, and strongly recommend this practice -- many VBs are possible, and a precise specification of \emph{which} VB was used is critical for reproducibility.  The figures in this section are intended to illustrate how an experimental paper might display \emph{real} data.  Each figure contains and compares multiple sub-figures, each with its own caption.  The sub-figure captions are in plain Roman type (\eg ``Figure 1(c)'') and illustrate how real data might be captioned.  Main figure captions are in bold (\eg ``\textbf{Figure 1:}''), and represent our own views.

For small, NISQ-scale processors, linear axes work well.  But displaying data from processors that can achieve width or depth $\gg1$ typically demands logarithmic axes.  Current state of the art processors can achieve depth $>$ 1000, but width at most $10-20$.  To display such capabilities, we generally mark the depth axis with integer powers of 2, but mark the width axis with integer powers of $1.2$ (rounded to an integer, duplicates removed). The rounding process results in an approximately linear scaling up to $\sim10$ qubits before transitioning to a more apparent logarithmic scale, as in Fig.~\ref{fig:example197}.

The easiest VBs to report are those with binary success criteria -- each shape (width/depth pair) either passes or fails.  In this situation, we indicate successes with a filled square.  Failed shapes can be indicated, if desired, with a hollow square (to distinguish ``test failed'' from ``test not performed'').  If a benchmark's success metric is real-valued, we place a solid marker at each tested shape and use shading to indicate the observed value of the success metric. 

Figure \ref{fig:two_qubits} illustrates how both binary and real-valued VB results can be displayed for a hypothetical two-qubit system.  Even a quick glance at Fig.~\ref{fig:two_qubits} reveals the most important result, which is that two-qubit ($w=2$) circuits perform significantly worse than single-qubit ($w=1$) circuits.  

The two-qubit examples demonstrate the core features of volumetric benchmarking plots, but become much more useful for larger processors.  The most relevant information from benchmarks with binary success criteria is contained in the \emph{Pareto frontier} that separates the regions of passed and failed tests.  The Pareto frontier is defined by the maximum depth, at each circuit width, for which the processor is reliable under the test metric. It bounds the feasible region.  

In most cases, the device's performance can be effectively conveyed \emph{just} by plotting the Pareto frontier, without showing the success/failure of every shape in the grid.  This provides simpler, cleaner plots and is especially useful for superposing and comparing results from multiple devices in a single figure. Figure \ref{fig:improve} illustrates this. Figures \ref{fig:example16} and \ref{fig:example197} provide further examples of how showing the Pareto frontier can either replace \emph{or} augment the basic grid display.

In many situations, prior information about the \emph{expected} performance of a processor is available.  This usually constitutes (or implies) a predictive model of the processor -- \eg one based on one- and two-qubit calibration experiments.  It can be extremely useful to compare experimental benchmarking data to the predictions of such a model.  We can do this by displaying both the empirical and predicted behavior on the same plot.  If they disagree, then the form of the discrepancy can highlight and help to identify emergent failure mechanisms that occur in sophisticated processors, such as crosstalk and non-Markovianity.

Figures \ref{fig:example16}, \ref{fig:example197}, and \ref{fig:qv} show how to do this.  In each of these figures, we display the experimental data as before using large squares, and add small squares to represent circuit shapes that the model predicted should pass. This technique can be extended to quantitative metrics as shown in Fig.~\ref{fig:example197}(b).  In that plot, we use shading to indicate quantitative performance, so the color of the large square indicates \emph{observed} performance, while the color of the small square indicates \emph{predicted} performance.  Discrepancies between the two are apparent by inspection. 

Figure \ref{fig:example197}(b) takes this one step further.  To highlight one advantage of plotting predictions and data together, we use red boxes to indicate regions on the plot where a specific discrepancy between observed and predicted performance suggests what might be going wrong.  In general, these discrepancies indicate that the model is insufficient to describe the dynamics of the device.  In particular, width deficiencies can indicate bad qubits, bad two-qubit gates, or the appearance of emergent crosstalk. Depth deficiencies can also arise from crosstalk, but for width-1 circuits they are indicative of non-Markovianity (or coherent errors if the benchmark being used is sensitive to them).

We conclude by showing how quantum volume relates to -- and can be shown within -- the VB visualization framework.  Obviously, making a rigorous connection to quantum volume as defined in Ref. \cite{Cross18} requires choosing and performing a specific benchmark (the one proposed in that work)!  But even for that benchmark, there are advantages to the richer reporting that we propose.  They are illustrated in Fig.~\ref{fig:qv}. In this example, the quantum volume experiments are explicitly highlighted along the diagonal, as is a region of implied success (those circuits with width \emph{and} depth less than or equal to the quantum volume).  This particular hypothetical device has a quantum volume of $2^8$, which is clear in the plot -- but it's capable of implementing significantly deeper circuits if restricted to slightly fewer qubits. As usual, logarithmic axes allow the display of a much wider range of circuit shapes, as shown in Fig.~\ref{fig:qv}(b). 

The sample volumetric benchmarking plots shown in the next pages were generated using Python code, which is provided as supplementary material.

\newpage

\addtocounter{figure}{1}
\begin{figure*}[htp]
\fbox{\begin{minipage}[t]{0.3\linewidth}
\centering
\normalfont\scriptsize
\includegraphics[width=\linewidth]{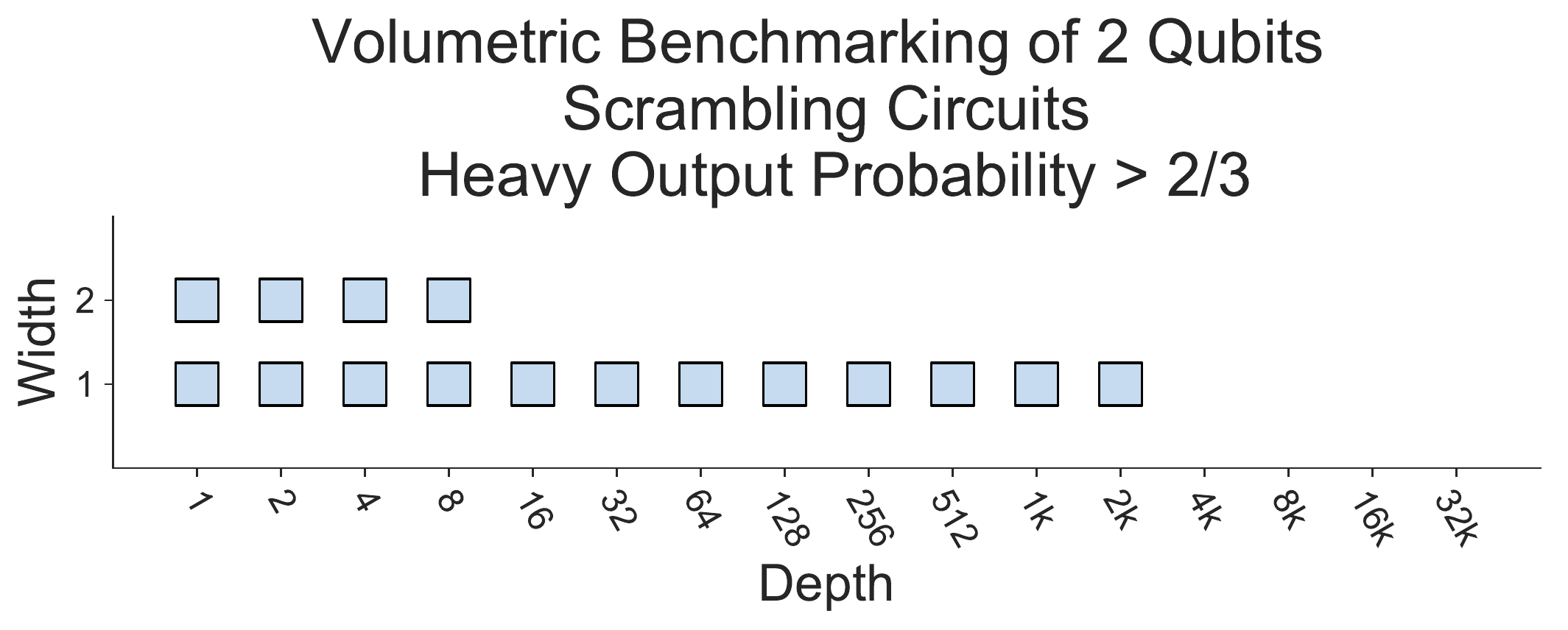}
\justify 
Figure~\thefigure(a). Two-qubit volumetric benchmarking using scrambling circuits. Circuit ensemble passes if the average heavy-output probability is greater than 2/3. These results indicate that while single-qubit circuits up to depth $\sim2000$ can be performed reliably, two-qubit circuits are limited to depth $\sim8$.
\end{minipage}\hspace{.4cm}
\begin{minipage}[t]{0.3\linewidth}
\centering
\normalfont\scriptsize
\includegraphics[width=\linewidth]{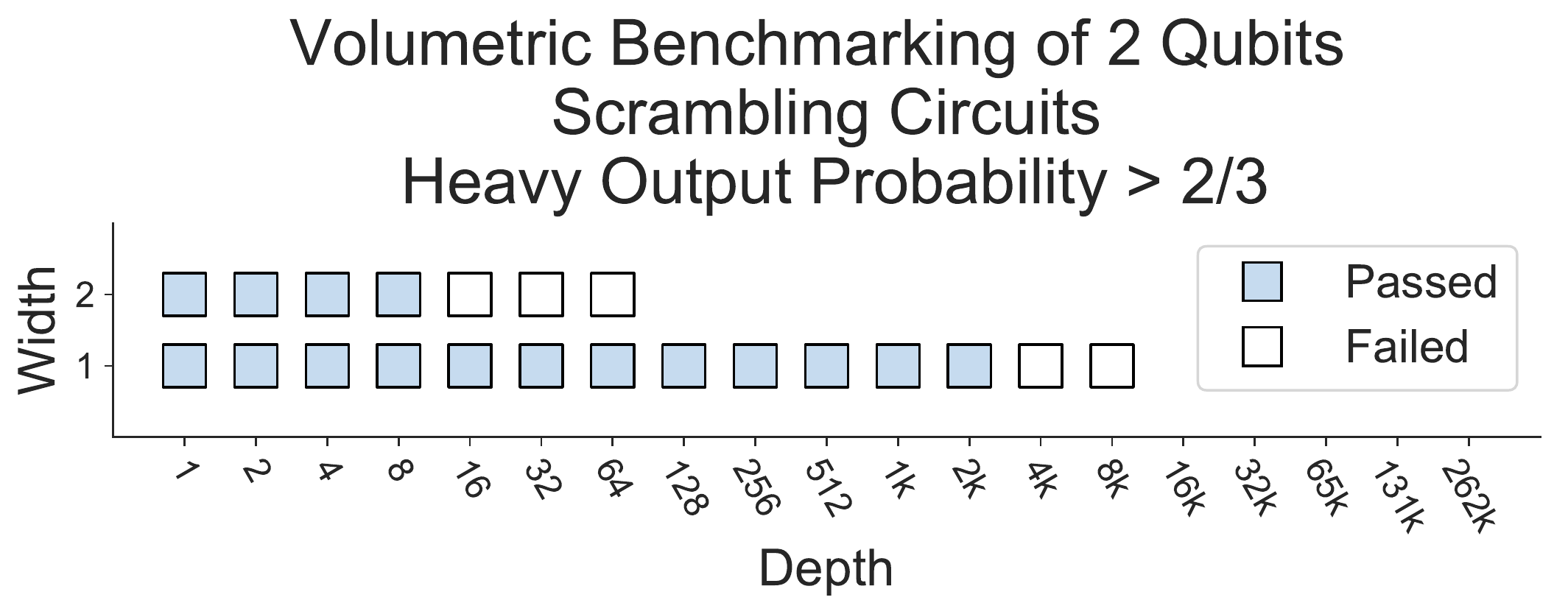}
\justify
Figure~\thefigure(b). Two-qubit volumetric benchmarking using scrambling circuits. Circuit ensemble passes if the average heavy-output probability is greater than 2/3. These results indicate that while single-qubit circuits up to depth $\sim2000$ can be performed reliably, two-qubit circuits are limited to depth $\sim8$.
\end{minipage}\hspace{.4cm}
\begin{minipage}[t]{0.3\linewidth}
\centering
\normalfont\scriptsize
\includegraphics[width=\linewidth]{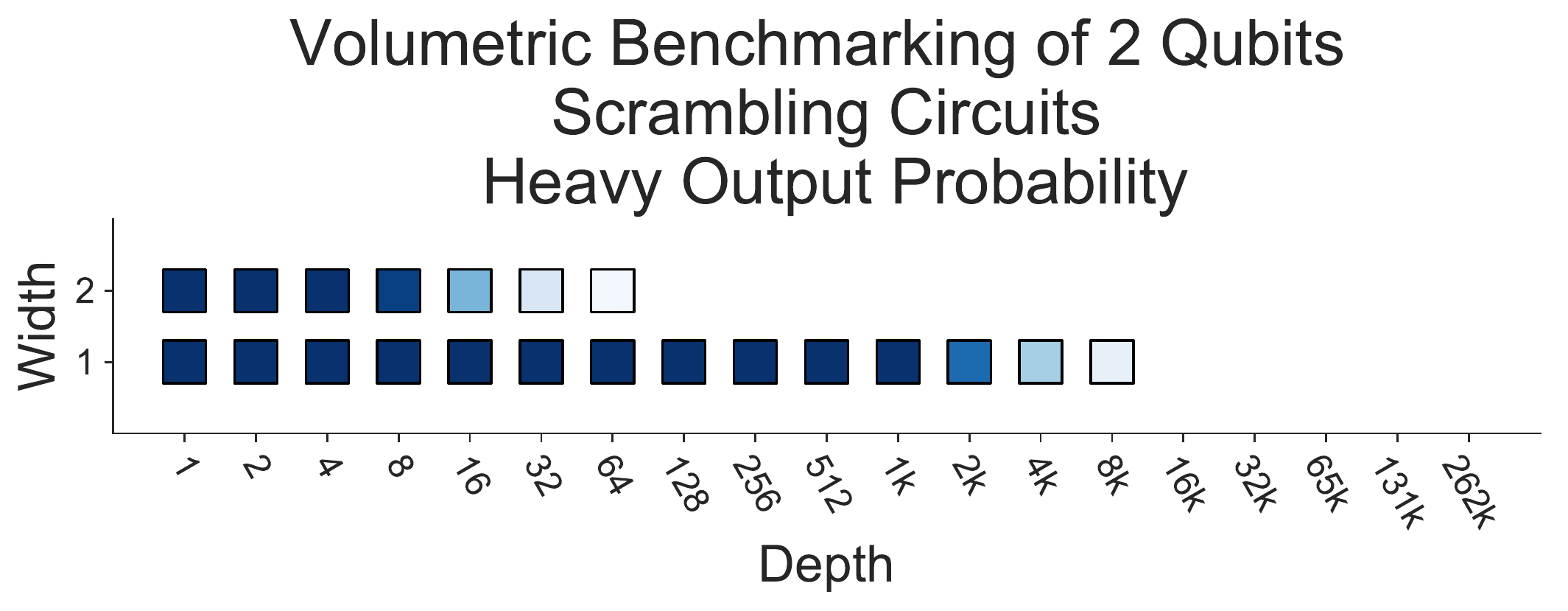}
\justify
Figure~\thefigure(c). Two-qubit volumetric benchmarking using scrambling circuits. Displayed is the heavy output probability for each width/depth pair tested. These results indicate that while single-qubit circuits up to depth $\sim2000$ can be performed reliably, two-qubit circuits are limited to depth $\sim8$.
\end{minipage}}
\addtocounter{figure}{-1}
\caption{Simple volumetric benchmarking plots for two qubits. The same data is displayed in each figure. (a) Only the width/depth pairs that passed the test are shown. (b) The width/depth pairs that failed are also indicated. (c) No specific pass/fail criterion is established. The average heavy outcome probability is indicated on a color scale for all width/depth pairs that were tested. \emph{This data is for illustration purposes only and does not come from a real device or rigorous simulation!}}
\label{fig:two_qubits}
\end{figure*}

\addtocounter{figure}{1}
\begin{figure*}[htp!]
\fbox{
\begin{minipage}[t]{0.3\linewidth}
\centering
\normalfont\scriptsize
\includegraphics[width=\linewidth]{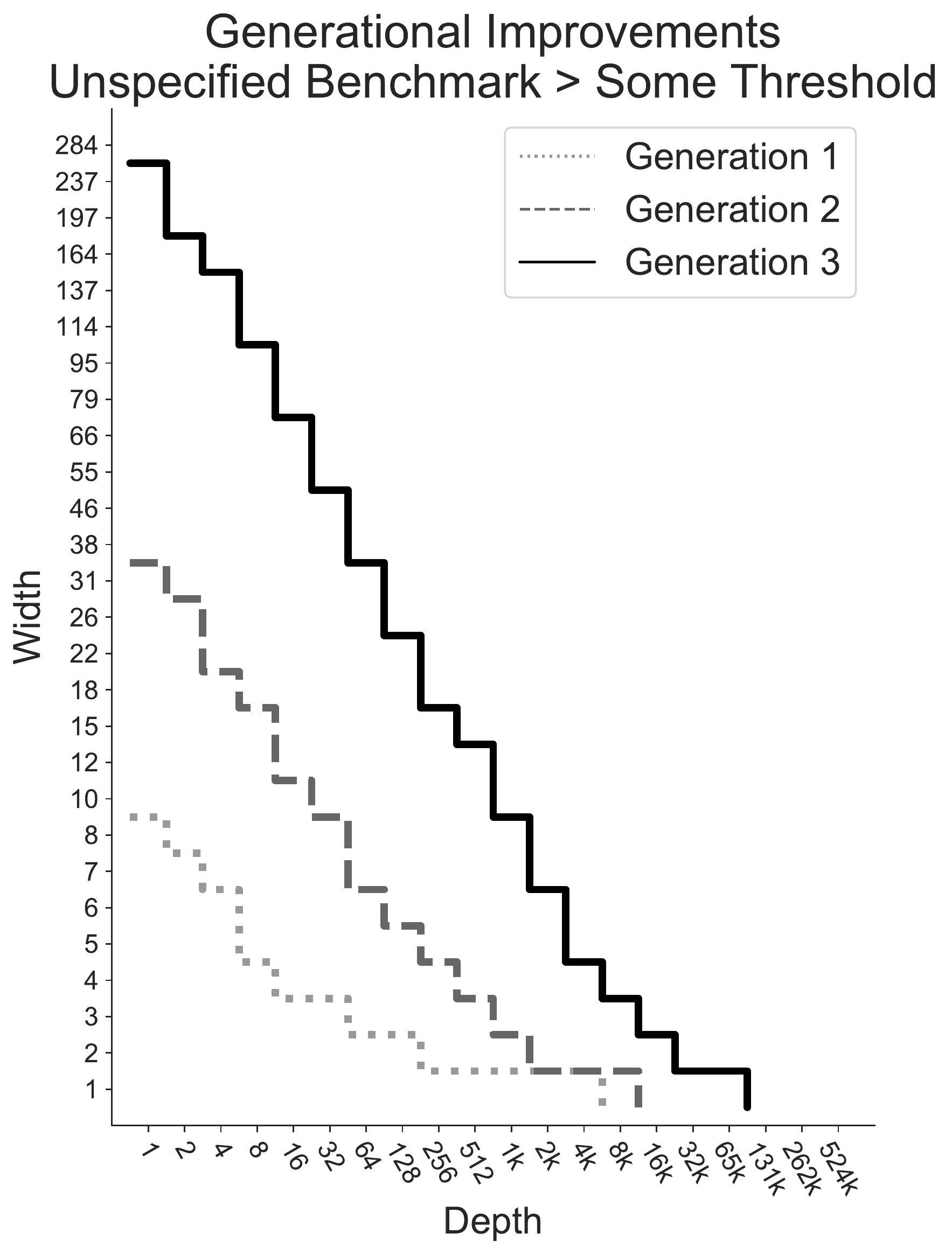}
\justify 
Figure~\thefigure(a). The Pareto frontier of an unspecified benchmark for three generations of quantum devices. Significant increases in both qubit performance and the number of available qubits can be seen between generations. 
\end{minipage}\hspace{.02\linewidth}
\begin{minipage}[t]{0.3\linewidth}
\centering
\normalfont\scriptsize
\includegraphics[width=\linewidth]{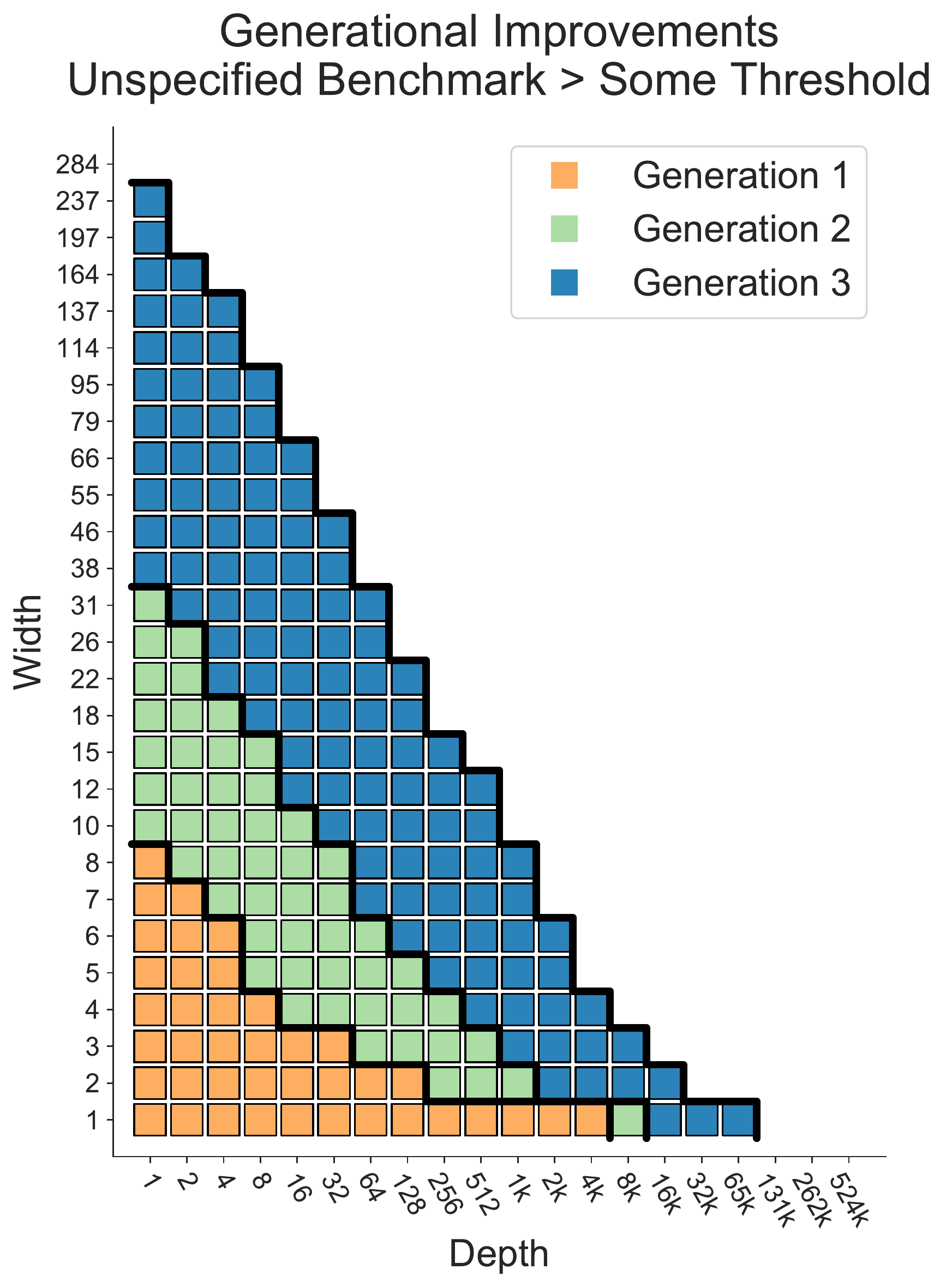}
\justify
Figure~\thefigure(b). The Pareto frontier of an unspecified benchmark for three generations of quantum devices. Significant increases in both qubit performance and the number of available qubits can be seen between generations. 
\end{minipage}\hspace{.02\linewidth}
\begin{minipage}[t]{0.3\linewidth}
\centering
\normalfont\scriptsize
\includegraphics[width=.95\linewidth]{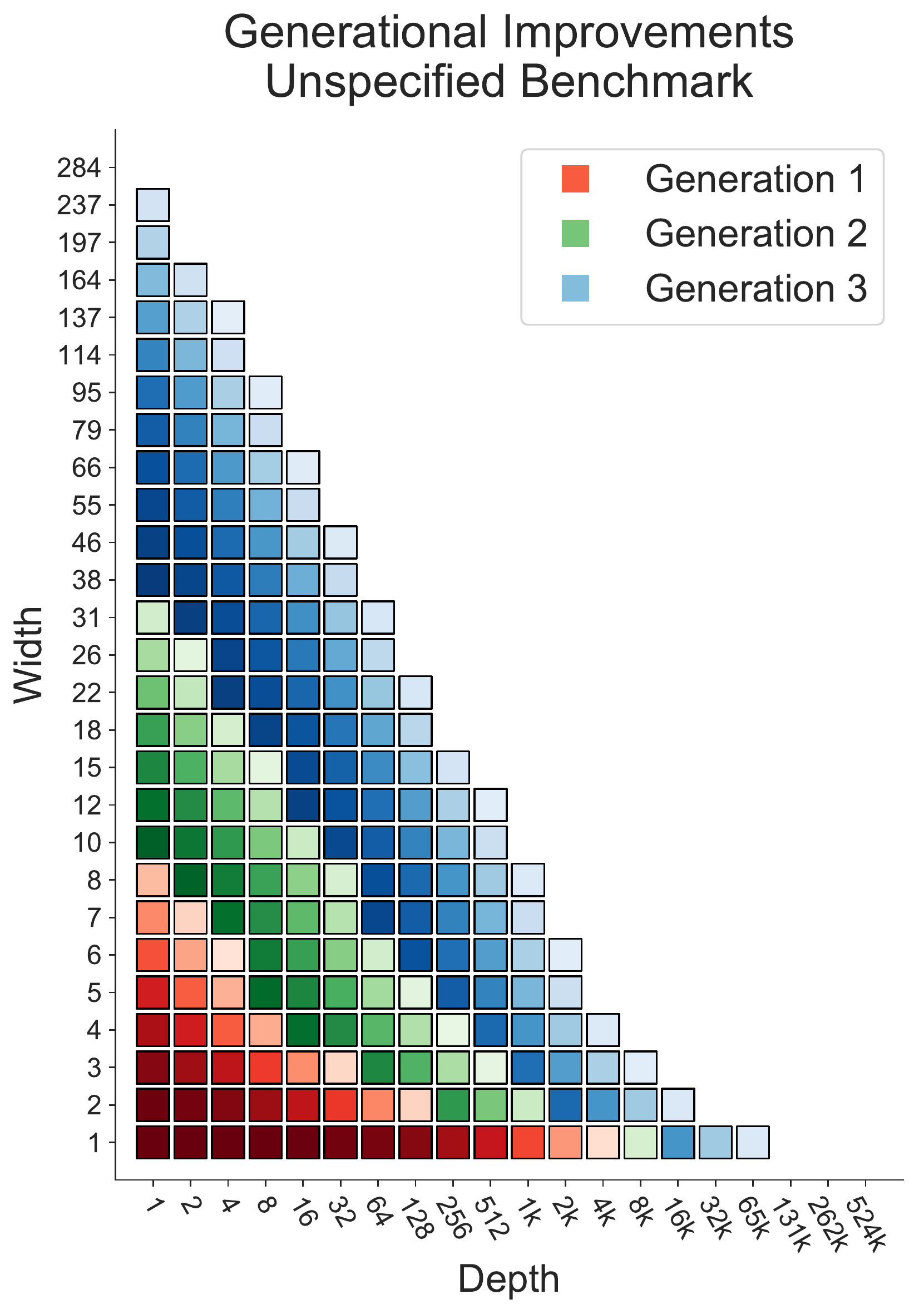}
\justify
Figure~\thefigure(c). The Pareto frontier of an unspecified benchmark for three generations of quantum devices. Significant increases in both qubit performance and the number of available qubits can be seen between generations. 
\end{minipage}}
\addtocounter{figure}{-1}
\caption{Volumetric benchmarking plots can be useful for comparing multiple devices, as long as the same benchmark and success metric are used.  All three of these figures display the same data. The improved slope in the Pareto frontier at later generations is a hallmark of reduced crosstalk. In (a), only the Pareto frontiers are displayed. In (b), the successful width/depth pairs are indicated as well, with earlier generation data plotted on top of the later generations. In (c), the successful experiments are shaded according to their value.  Figure (c) is rather cluttered, however, and we don't recommend this approach.  \emph{This data is for illustration purposes only and does not come from a real device or rigorous simulation!}}
\label{fig:improve}
\end{figure*}

\addtocounter{figure}{1}
\begin{figure*}[htp!]
\fbox{\begin{minipage}[t]{0.48\linewidth}
\centering
\normalfont\scriptsize
\includegraphics[width=0.6\linewidth]{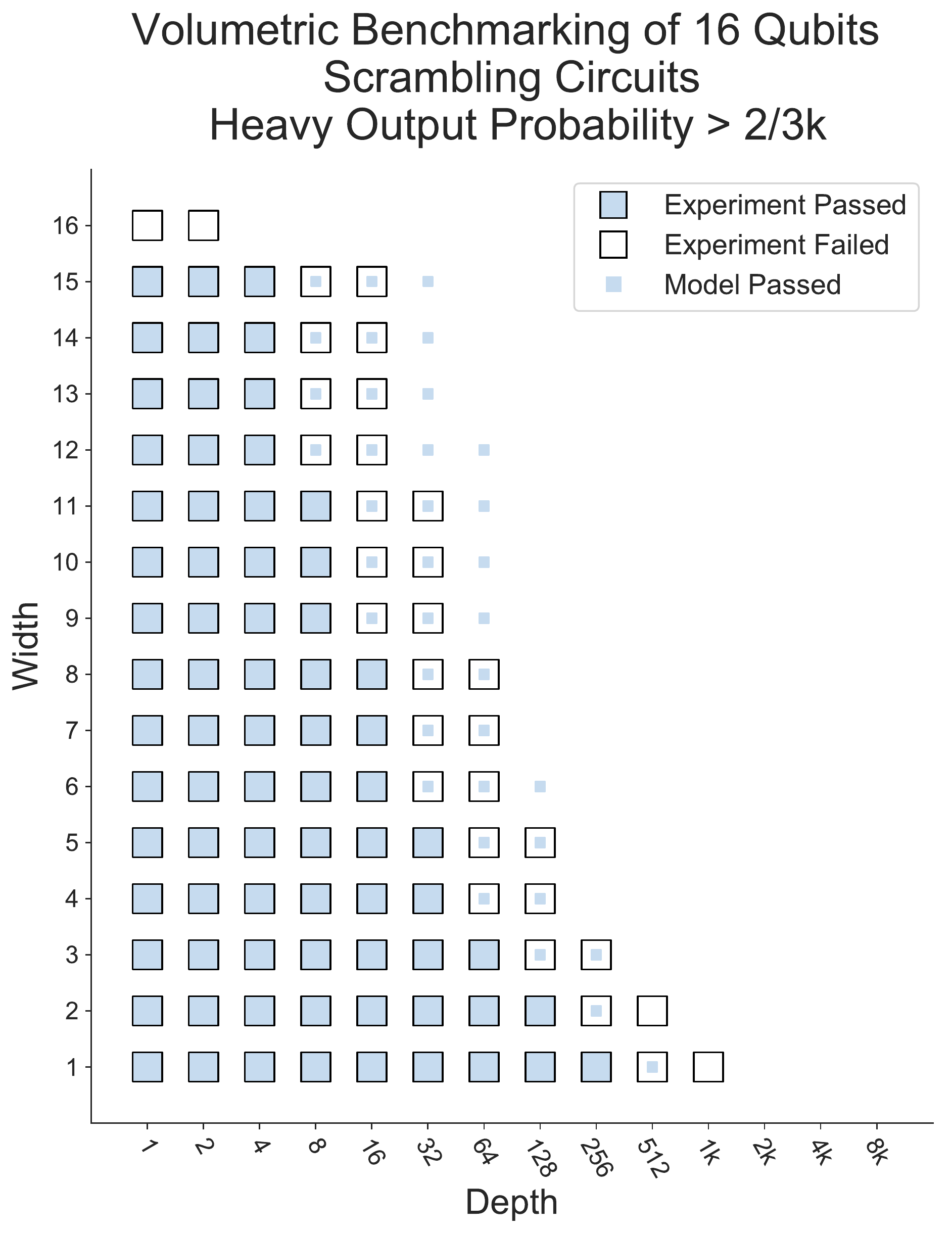}
\justify 
Figure~\thefigure(a). 16-qubit volumetric benchmarking using scrambling circuits. Pass/fail criterion: heavy-output probability $>2/3$. The model was constructed using one- and two-qubit experiments. Qubit 16 was defective.
\end{minipage}\hspace{.5cm}
\begin{minipage}[t]{0.48\linewidth}
\centering
\normalfont\scriptsize
\includegraphics[width=.6\linewidth]{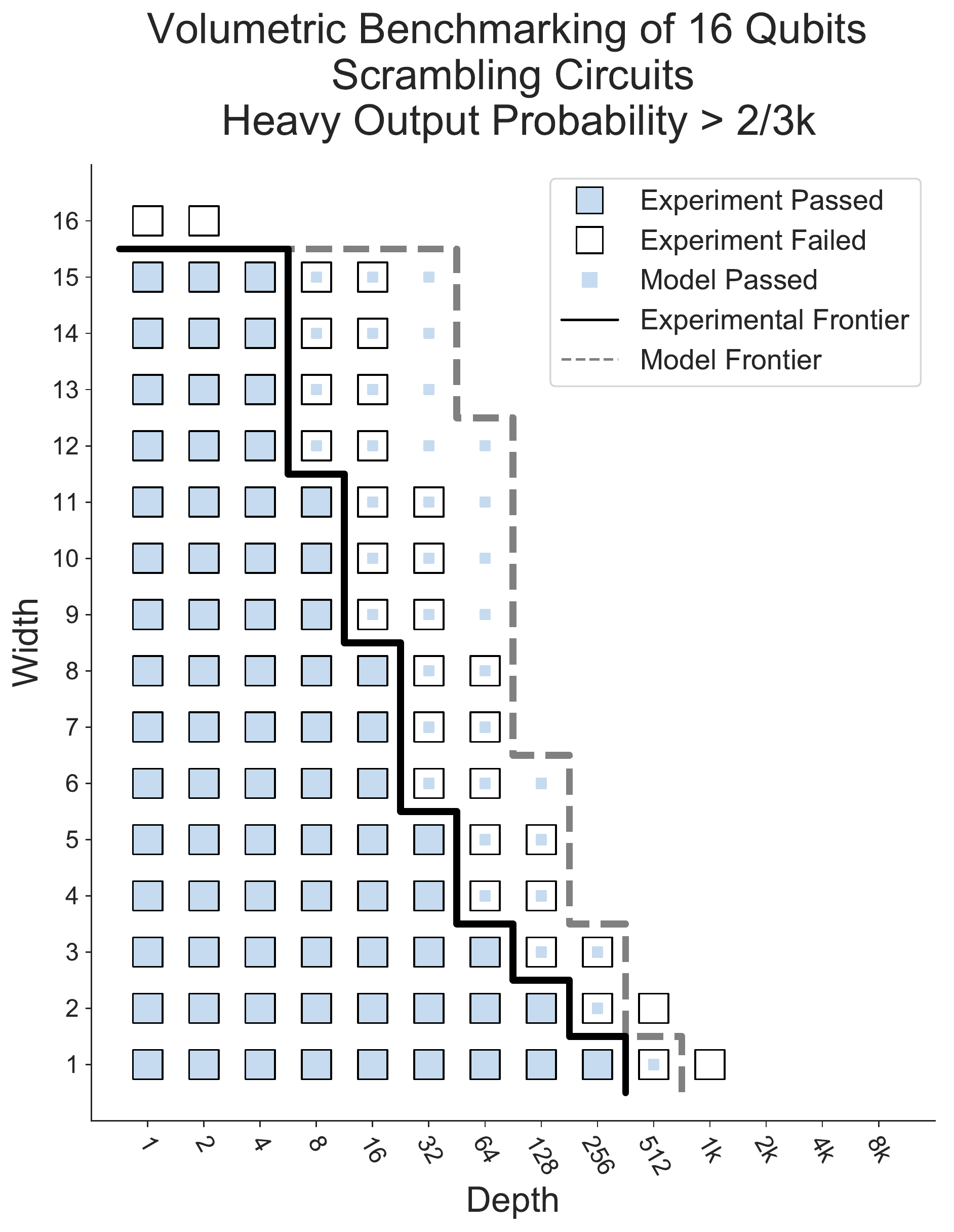}
\justify
Figure~\thefigure(b). 16-qubit volumetric benchmarking using scrambling circuits. Pass/fail criterion: heavy-output probability $>2/3$. The model was constructed using one- and two-qubit experiments. Qubit 16 was defective.
\end{minipage}}
\addtocounter{figure}{-1}
\caption{ Including the Pareto frontier can improve the readability of the figure. Deviation of the data and model predictions can indicate unexpected failure mechanisms, such as non-Markovianity and crosstalk. \emph{This data is for illustration purposes only and does not come from a real device or rigorous simulation!}}
\label{fig:example16}
\end{figure*}

\addtocounter{figure}{1}
\begin{figure*}[htp!]
\fbox{\begin{minipage}[t]{0.48\linewidth}
\centering
\normalfont\scriptsize
\includegraphics[width=0.6\linewidth]{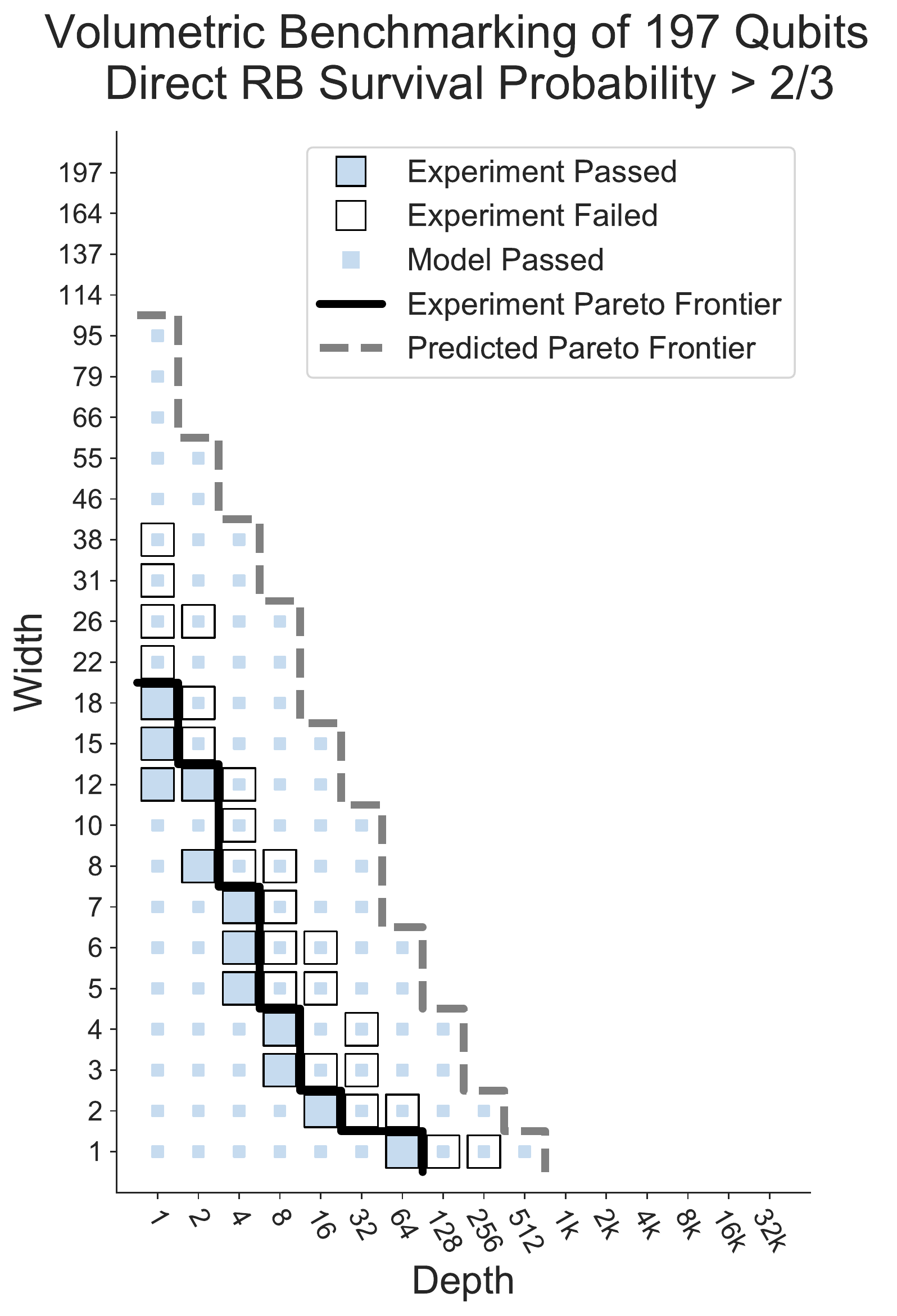}
\justify 
Figure~\thefigure(a). Performance of hypothetical 197-qubit device benchmarked with direct randomized benchmarking (DRB). Width/depth pair passes test if the average DRB survival rate is $> 2/3$. Model predictions are based on a depolarizing noise model with parameters obtained using one- and two-qubit randomized benchmarking.  Only experiments near the Pareto frontier were performed. 
\end{minipage}\hspace{.5cm}
\begin{minipage}[t]{0.48\linewidth}
\centering
\normalfont\scriptsize
\includegraphics[width=.6\linewidth]{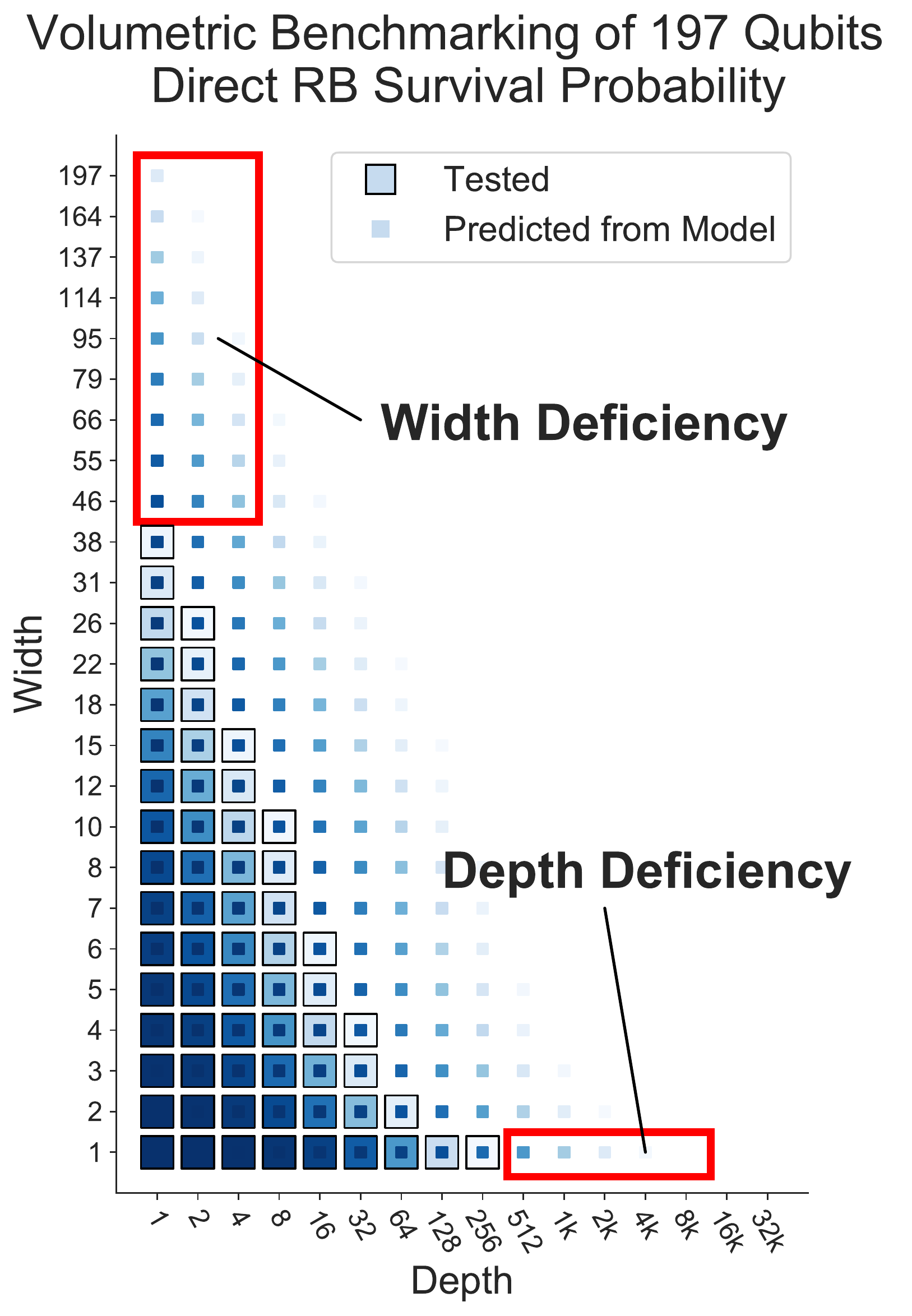}
\justify
Figure~\thefigure(b). Performance of hypothetical 197-qubit device benchmarked with direct randomized benchmarking (DRB).  Model predictions are based on a depolarizing noise model with parameters obtained using one- and two-qubit randomized benchmarking.
\end{minipage}}
\addtocounter{figure}{-1}
\caption{Using VBs to compare experimental data against predictions of a model.  Discrepancies between the experimental performance and the model's predictions can give insight into unexpected failure mechanisms.  For instance, a width deficiency can indicate the presence of crosstalk or some number of bad qubits, while a depth deficiency could indicate that some of the gates have non-Markovian or coherent errors. \emph{This data is for illustration purposes only and does not come from a real device or rigorous simulation!}}
\label{fig:example197}
\end{figure*}

\addtocounter{figure}{1}
\begin{figure*}[htp!]
\fbox{
\begin{minipage}[t]{\linewidth}
\centering
\normalfont\scriptsize
\includegraphics[width=.95\linewidth]{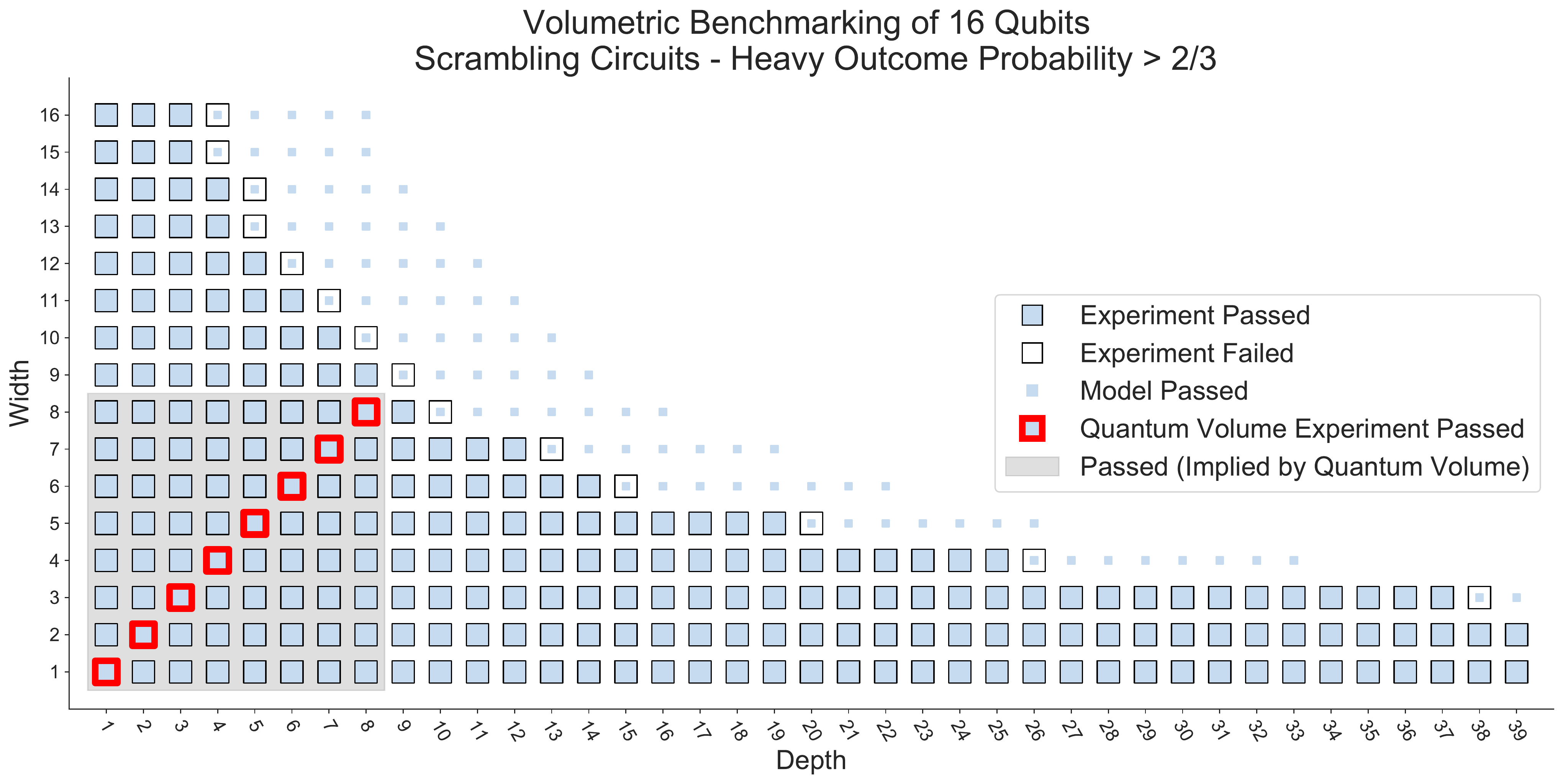}
\justify 
Figure~\thefigure(a). Volumetric benchmarking of a 16 qubit device using scrambling circuits. If at least $2/3$ of the measurement results are heavy for a given width/depth pair, then the pair passes the test and is marked with a large, solid blue box. Using linear axes, the quantum volume experiments appear along the diagonal and are outlined with heavy, red lines. For this example, $\log_2(V_Q)=8$. It is expected that scrambling circuits with both width and depth less than or equal to the quantum volume should succeed, and we highlight these with a gray background. 
\end{minipage}}\vspace{.5cm}
\fbox{\begin{minipage}[t]{\linewidth}
\centering
\normalfont\scriptsize
\includegraphics[width=.6\linewidth]{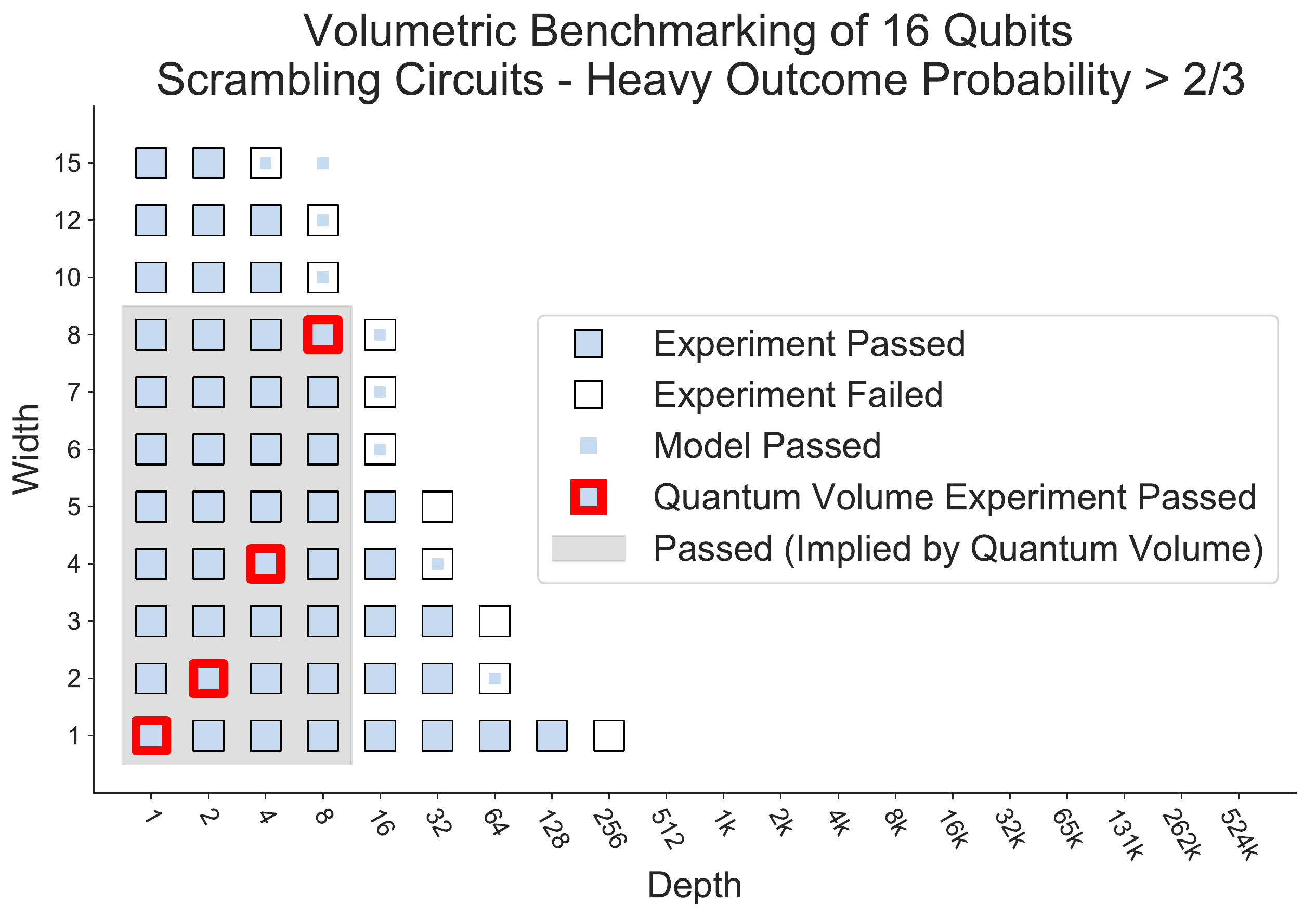}
\justify
Figure~\thefigure(b). Volumetric benchmarking of a 16 qubit device using scrambling circuits. If at least $2/3$ of the measurement results are heavy for a given width/depth pair, then the pair passes the test and is marked with a large, solid blue box. Using logarithmic axes, the quantum volume experiments appear along a curved line and are outlined with heavy, red lines. For this example, $\log_2(V_Q)=8$. It is expected that scrambling circuits with both width and depth less than or equal to $\log_2(V_Q)$ should succeed, and we highlight these with a gray background.
\end{minipage}}
\addtocounter{figure}{-1}
\caption{Logarithmic axes allow more data (more circuit shapes) to be compressed into the same space. In Fig.~\ref{fig:qv}(b), the depths are powers of 2 and the widths are powers of 1.2 (rounded and with duplicates removed). This width scaling was chosen to deal with the relatively small size of contemporary devices.  \emph{This data is for illustration purposes only and does not come from a real device or rigorous simulation!}}
\label{fig:qv}
\end{figure*}

\clearpage

\begin{acknowledgments}
The authors thank Kenneth Rudinger for his helpful -- and comprehensive -- comments on a late draft of this manuscript and Stephen Bartlett for graciously providing an opportunity to write this paper (at great personal cost). All statements of fact, subjective views, opinions, or conclusions expressed herein are strictly those of the authors; they do not represent the official views or policies of the Department of Energy or the U.S. Government. Sandia National Laboratories is a multi-mission laboratory managed and operated by National Technology \& Engineering Solutions of Sandia, LLC, a wholly owned subsidiary of Honeywell International Inc., for the U.S. Department of Energy's National Nuclear Security Administration under contract {DE-NA0003525}. This material was funded in part by the U.S. Department of Energy, Office of Science, Office of Advanced Scientific Computing Research Quantum Testbed Program.
\end{acknowledgments}

\bibliographystyle{unsrtnat}
\bibliography{VolumetricBenchmarks}

\end{document}